\documentclass[11pt]{amsart}
\usepackage{geometry}                % See geometry.pdf to learn the layout options. There are lots.
\usepackage{graphicx}
\usepackage{amssymb}
\usepackage{epstopdf}
\usepackage{array}
\usepackage{natbib}
\usepackage{hyperref}

\usepackage{amsmath}
\usepackage{amsthm}
\usepackage{verbatim}

\usepackage{tikz}
\usetikzlibrary{scopes}
\usetikzlibrary{decorations.pathmorphing}
\usetikzlibrary{arrows}
\pgfdeclarelayer{nodelayer}
\pgfdeclarelayer{edgelayer}
\pgfsetlayers{edgelayer,nodelayer,main}
\tikzstyle{plain}=[rectangle,fill=none,draw=white,scale=1.0,inner sep=1.8pt] % circle
\tikzstyle{line}=[draw=black]
\tikzstyle{arrow}=[draw=black,arrows=-latex]
\tikzstyle{level}=[circle,draw=black,thick]
\tikzstyle{blank}=[circle,draw=white, inner sep=0.1pt]
\tikzstyle{tensor}=[circle,draw=white, inner sep=4.0pt]

\title{Fast functional integrals with application to differential equations models}
\author{John Tillinghast}

\begin{document}

%\footnote{This work was begun at the Johns Hopkins Bloomberg School of
%Public Health under Training Grant T32AG000247 (epidemiology and biostatistics of
%aging).}

\maketitle

\section{Abstract}
%\subsection{}

A new method is introduced which uses higher-order Laplace approximation to evaluate functional integrals much faster than existing methods. An implementation in MATLAB is called SLAM-FIT (Sparse Laplace Approximation Method for Functional Integration on Time) or simply SLAM. In this paper SLAM is applied to estimate parameters of mixed models that require functional integration.
It is compared with two more general packages which can be used to do functional integration. One is Stan, a recent and very general package for integrating and estimating using hybrid Monte Carlo. The other is INLA, a recent R package which uses Laplace approximations for Gaussian Markov random fields. In both cases it is able to get near-identical or equivalent results in less time for moderately sized data sets. The fundamental speed advantage of the algorithm may be greater than it appears, because SLAM is running in pure MATLAB while the other two packages use optimized compiled code.

%SLAM is compared with Stan on an infectious disease data set modeled with a stochastic differential equation. Here SLAM gets equivalent estimates at much greater speed. Then SLAM is compared with INLA for repeats of an INLA demonstration data set, Tokyo rainfall, which is a GMRF on the line. Estimates are again equivalent to within a fraction of the confidence interval. When both packages are using higher-order terms, SLAM overtakes INLA after a few repeats, and can analyze a larger data set without being stopped by memory limits.

Keywords: compartment models, predator-prey, SIR model, path integral, functional integral, Laplace approximation, saddlepoint method.

\section{Introduction}

This paper has two purposes: to introduce a new, much faster computational method for functional integration, and to show its usefulness in estimating parameters of differential equation models. For a time-dependent random process, a functional integral can be thought of as the integral over all possible values at all times over the relevant time interval ~\citep{Dirac, FH, Schulman, Kleinert}. Technically they are integrals over a function space such as a space of Brownian motion paths. They have been used for at least eighty years in many branches of science and applied mathematics including quantum and statistical mechanics, polymer science, probability, and finance. Typically they are used to compute the chance of a system evolving from one state to a different one: any in-between path is possible, but the probability is given by the integral over the possible paths.
This makes them a natural tool to use for analyzing systems which are imperfectly modeled by ordinary differential equations. In life science there are predator-prey models ~\citep{MathModels}, infectious disease models ~\citep{SIR1927, Anderson}, pharmacokinetics models ~\citep{Gelman1996}, and many others. Most of macroeconomics depends on differential equation-based models such as the Solow growth model or the ISLM model ~\citep{Romer}.  These models can be made more realistic by considering noise or stochastic behavior.

A major goal of differential equations models is to estimate system parameters, such as how infectious a virus is, or reaction rates for chemicals and enzymes. Traditional approaches involve simplified models on transformed variables ~\citep{LWB} or optimizing some measure of data fit over exact solutions to the ODEs ~\citep{Anderson}. 
Recent methods use a generalized smoothing approach. For a finite-dimensional basis, such as splines, it is possible to approximately fit both the differential equations and the data ~\citep{RH, Tempering}. As with smoothing, the trade-off between data fit and ODE fit can be chosen by cross-validation or by a mixed-model approach.

In this paper, we use a very different approach: we discretize time, and model the underlying true values as a first-order Markov process. Then we assume that the data were generated with a distribution depending on the true value. Not all of the time points have data; in fact, it may be important to add in-between time points, just as they would be needed for a finite difference method. Then we define a marginalized pseudolikelihood of the system parameters as an integral over the possible values which the variables might have taken at the time points. This approximates a functional integral over a continuous random process.

%If we can integrate this over all possible realizations, we can define a marginalized likelihood of the system parameters, given the data. This is a functional integral which must be maximized in order to estimate the parameters.

There are both analytic and computational methods for evaluating functional integrals. Analytic methods require a tractable integrand or a good approximation to one. Very clever schemes have been painstakingly developed to transform functional integrals and make them tractable ~\citep{Kleinert, Schulman}. Far more problems have so far required time-consuming Monte Carlo methods.

Laplace and saddlepoint approximations have also been used to approximate analytic functional integrals when possible. But this appears to be the first time that the approximations have been used for numerical evaluation. One reason for this may be that some of the higher-order terms, involving third- and fourth-order tensors, are difficult to evaluate efficiently.

This paper introduces a way to compute the higher-order terms quickly for numerical functional integrals. This requires computing a critical path (a unique most likely realization) and expanding the log-likelihood around it. The log-likelihood is a sum of interactions between neighboring time points. Consequently the relevant Hessian and the tensors of third- and fourth-order derivatives are sparse with a simple block structure: all entries corresponding to non-neighboring time points are zero. By using this sparsity structure, it is possible to calculate the higher-order terms in $O(np^4)$ time, where $n$ is the number of time points and $p$ is the number of variables in the system.

A variance-stabilizing transformation is needed when the data times are separated by many in-between time steps. In general it makes the integrals more accurate than when untransformed. SLAM can be used with the basic Laplace term alone, or with the higher-order terms. Higher-order terms make the integrals more accurate, but for the examples tested, the parameter estimates are close to those found using the basic Laplace approximation. 
%[For the final article you will want to display this]

To evaluate speed and accuracy, SLAM was compared with two popular systems for mixed models. Results for an infectious disease data set are equivalent to results from Stan ~\citep{STAN}, the leading general-purpose Monte Carlo system, but SLAM is much faster. There is also a well-implemented Laplace-based method, INLA ~\citep{INLA, INLA-URL}, designed for hierarchical models and Gaussian Markov random fields. 

For random fields, INLA takes advantage of sparsity, but the higher-order terms still take quadratic time. The special approximation introduced here can only be applied to GMRFs on a line, not for other type covered by INLA. To compare INLA and SLAM, we used an INLA demonstration data set (Tokyo rainfall by day of year). For repeats of the Tokyo data, estimates are well within the CI for the two methods. INLA (with full higher-order terms) is faster at 1100 data points but SLAM is faster beyond that. INLA using simplified higher-order terms remains somewhat faster up to the largest set tested (about 15000 data points).

%\subsection{Questions for reviewer}
%
%Does it make sense to include the numbers of data points? This is the intro after all.
%SLAM or FFIT?
%Should I mention $O(np^4)$ in the abstract as well as the intro?
%The results about accuracy of the integrals, importance sampling, etc. are not shown in this version of the paper. Is it a mistake to mention them here? The basic Laplace approximation isn't used in the INLA example because the internal computations should be almost identical to INLA's.

\section{METHODS}

\subsection{Markov process setup}

The model has two stages: a Markov process for the true values, and on top of that, measurements with noise. We assume that the Markov and measurement likelihoods depend on parameters $\boldsymbol{\theta}$. We show how to define a marginalized likelihood for $\boldsymbol{\theta}$. Maximizing the marginalized likelihood gives an estimate of $\boldsymbol{\theta}$.

In this paper we will use Latin indexes (``i") to denote time points and matrix blocks. We use Greek indexes (``$\alpha$") to denote individual vector and matrix entries. Each time point has as one entry for every variable. This becomes important when analyzing the blocking structure. We also use the summation convention for duplicated indexes (e.g. $v_\gamma = A_{\alpha \beta} T_{\alpha \beta \gamma}$ means $v_\gamma = \sum_{\alpha, \beta} A_{\alpha \beta} T_{\alpha \beta \gamma}$.)

The Markov process has a vector value at each time point, denoted by $\mathbf{y}_i$ at step $i$ (time $t_{i}$). (We use $\mathbf{y}$ without an index to mean the entire realization, i.e.,  the concatenation of the $\mathbf{y}_i$.) The system also has parameters $\boldsymbol{\theta}$ that involved in the Markov process or the measurement error. From the Markov process, at time $t_{i+1}$ there is a conditional probability density for $\mathbf{y}_{i+1}$, given the value of $\mathbf{y}_i$:

\begin{equation*}
f_{trans} \left(\mathbf{y}_{i+1} | \mathbf{y}_i, \boldsymbol{\theta} \right) 
= exp \big(  -\ell_{trans} \left(\mathbf{y}_{i+1} | \mathbf{y}_i, \boldsymbol{\theta} \right) \big)
\end{equation*}
so given the initial values $\mathbf{y}_1$,  we have the overall conditional density for a realization $\mathbf{y}$ at all time points:

\begin{equation*}
f_{dyn}  \left( \mathbf{y} |  \mathbf{y}_1, \boldsymbol{\theta} \right)  
  = \prod_{i=1}^{n-1}  f_{trans}\left(\mathbf{y}_{i+1} | \mathbf{y}_i, \boldsymbol{\theta} \right) \\
\end{equation*}
If we have a prior $\pi(\mathbf{y}_1)$ we can define the probability of a realization:

\begin{equation*}
f_{dyn}  ( \mathbf{y} | \boldsymbol{\theta} ) = \pi(\mathbf{y}_1) f_{dyn} \left( \mathbf{y}  |  \mathbf{y}_1, \boldsymbol{\theta} \right)  
\end{equation*}

Some of the time points have data. Let $I_{data}$ be the set of their indexes. Call the data values $\mathbf{y}_i^\ast$ for $i \in I_{data}$. We assume that the data are independently generated with a probability density depending on the true value and parameters:

\begin{equation*}
f_{data} \left(\mathbf{y}_i ^ {\ast} | \mathbf{y}_i, \boldsymbol{\theta} \right)
   = exp \big(  -\ell_{data}  \left(\mathbf{y}_i ^ {\ast} | \mathbf{y}_i, \boldsymbol{\theta} \right) \big)
\end{equation*}

Depending on the system, the error could be measurement error from an instrument, or from sampling error, or from some other source.  The overall density of the given realization and data is

\begin{equation*}
f(\mathbf{y, y^\ast} | \boldsymbol{\theta}) = \pi(\mathbf{y_1}) \prod_{i=1}^{n-1} f_{dyn} \left(\mathbf{y}_{i+1} | \mathbf{y}_i, \boldsymbol{\theta} \right) \prod_{i \in I_{data}} f_{data}(\mathbf{y}_i^\ast | \mathbf{y}_{i}, \boldsymbol{\theta} )
\end{equation*}
This means that the pdf of $\mathbf{y}^\ast$ given $\boldsymbol{\theta}$ is given by

\begin{align*}
f \left( \mathbf{y}^\ast | \boldsymbol{\theta}  \right) &= \int_{\mathbb{R}^N} \pi(\mathbf{y_1}) \prod_{i=1}^{n-1} f_{dyn} \left(\mathbf{y}_{i+1} | \mathbf{y}_i, \boldsymbol{\theta} \right) \prod_{i \in I_{data}} f_{data}(\mathbf{y}_i^\ast | \mathbf{y}_{i}, \boldsymbol{\theta} )
d^N \mathbf{y} \nonumber \\
&= \exp \left(   log \ \pi(\mathbf{y_1})   -\sum_{i \in I_{data}} \ell_{data} \left(\mathbf{y}_i^\ast | \mathbf{y}_i, \boldsymbol{\theta} \right) -\sum_{i=1}^{n-1}  \ell_{trans} \left(\mathbf{y}_{i+1} | \mathbf{y}_i, \boldsymbol{\theta} \right) \right)
\end{align*}

Given $\pi(\mathbf{y_1})$ it is possible to define a marginal likelihood for $\boldsymbol{\theta}$ by integrating $f \left( \mathbf{y}^\ast | \boldsymbol{\theta}  \right)$ over the latent true values. But in general, we don't know $\pi(\mathbf{y_1})$, and we use the improper prior $\pi(\mathbf{y_1}) = 1$. The integral is finite because of the multiplication by $f_{data}$ which can be thought of as a prior on the data points. This gives us the marginalized pseudolikelihood

\begin{align*}
M \left( \boldsymbol{\theta}  \right) &= \int_{\mathbb{R}^N} \prod_{i=1}^{n-1} f_{dyn} \left(\mathbf{y}_{i+1} | \mathbf{y}_i, \boldsymbol{\theta} \right) \prod_{i \in I_{data}} f_{data}(\mathbf{y}_i^\ast | \mathbf{y}_{i}, \boldsymbol{\theta} ) d^N \mathbf{y} \\
&= \int_{\mathbb{R}^N} e^{-\ell(\mathbf{y})}
d^N \mathbf{y} \nonumber
\end{align*}

where

\begin{equation*}
\label{ell}
\ell(\mathbf{y}) = \sum_{i \in I_{data}} \ell_{data} (\mathbf{y}_i^\ast | \mathbf{y}_{i}, \boldsymbol{\theta} ) + \sum_1^{N-1} \ell_{trans} \left(\mathbf{y}_{i+1} | \mathbf{y}_i, \boldsymbol{\theta} \right).
\end{equation*}

This marginalized pseudolikelihood is what we will maximize to estimate $\boldsymbol{\theta}$.

%\subsection{Questions for reviewer}
%
%
%Is the terminology correct here? Should this be called a pseudo likelihood, or an empirical Bayes approach, or something else?
%
%Is there a better argument for not having to use the prior?
%
%Is there a good way to describe the data probability as a prior?

%\subsection{Specifics for numerical SDEs}
 
%\subsection{Importance Sampling}

%Back to what's in Arvix version

\subsection {Laplace approximation (basic and higher-order)}

In this subsection, we look closely at how to get a good approximation for $M$. 
In addition, assume that $\ell (\mathbf{y})$ has a single peak at $\hat{\mathbf{y}}$.
Then we can expand around $\hat{\mathbf{y}}$:

\begin{align*}
\mathbf{y} &= \hat{\mathbf{y}} + \boldsymbol{\varepsilon} \nonumber \\
M &= \int_{\mathbb{R}^N} e^{ - \ell \left (\mathbf{y}\right)} d^N \mathbf{y} \\
&= \int_{\mathbb{R}^N} e^{ - \ell \left( \hat{\mathbf{y}} + \boldsymbol{\varepsilon} \right)} d^N \boldsymbol{\varepsilon} \nonumber \\
%&= \int_{\mathbb{R}^N} exp \left(- \ell \left(\hat{\mathbf{y}} + \varepsilon\right)\right) d^N \varepsilon \nonumber \\
&= \int_{\mathbb{R}^N} \exp \left( - \ell \left(\hat{\mathbf{y}}\right) 
- \frac{1}{2} \ell_{\alpha \beta}^{\left(2\right)} \left(\hat{\mathbf{y}}\right) \varepsilon_\alpha \varepsilon_\beta 
- \frac{1}{3!}  \ell_{\alpha \beta \gamma}^{\left(3\right)} \left(\hat{\mathbf{y}}\right) \varepsilon_\alpha \varepsilon_\beta \varepsilon_\gamma 
- \frac{1}{4!} \ell_{\alpha \beta \gamma \lambda}^{\left(4\right)} \left(\hat{\mathbf{y}}\right)
- ... \right) d^N \boldsymbol{\varepsilon} \nonumber \\
&= e^{ - \ell \left(\hat{\mathbf{y}}\right)} \int_{\mathbb{R}^N}  e^{- \frac{1}{2} H_{\alpha \beta} \varepsilon_\alpha \varepsilon_\beta} \exp \left( -\frac{1}{3!}  T_{\alpha \beta \gamma} \varepsilon_\alpha \varepsilon_\beta \varepsilon_\gamma -\frac{1}{4!}  F_{\alpha \beta \gamma \lambda} \varepsilon_\alpha \varepsilon_\beta \varepsilon_\gamma \varepsilon_\lambda - ... \right) d^N \boldsymbol{\varepsilon} \nonumber
\end{align*}
where $\mathbf{H}$ is the Hessian of $\ell (\mathbf{y})$ at $\hat{\mathbf{y}}$, and $\mathbf{T}$ and $\mathbf{F}$ are the tensors of third and fourth derivatives at $\hat{\mathbf{y}}$.

We can think of $\boldsymbol{\varepsilon}$ as a Gaussian random variable, with precision matrix $\mathbf{H}$, and this integral is the expectation of a function of $\boldsymbol{\varepsilon}$:

\begin{equation*}
M =
  e^{- \ell \left(\hat{\mathbf{y}}\right)} (2 \pi)^{N/2} \vert \mathbf{H} \vert^{-1/2}
  E \left\{
\exp \left( -\frac{1}{3!}  T_{\alpha \beta \gamma} \varepsilon_\alpha \varepsilon_\beta \varepsilon_\gamma -\frac{1}{4!}  F_{\alpha \beta \gamma \lambda} \varepsilon_\alpha \varepsilon_\beta \varepsilon_\gamma \varepsilon_\lambda - ... \right)
  \right\} 
\end {equation*}

The expectation can be expanded and then approximated with an asymptotic series.

\begin{align*}
M &\propto E \left\{   \sum_{r=0}^{\infty} 
      \left(-\right)^r \frac{1}{r!} 
      \left( \frac{1}{3!}  T_{\alpha \beta \gamma} \varepsilon_\alpha \varepsilon_\beta \varepsilon_\gamma + 
          \frac{1}{4!}  F_{\alpha \beta \gamma \lambda} \varepsilon_\alpha \varepsilon_\beta \varepsilon_\gamma \varepsilon_\lambda + ... 
      \right)^r  \right\} \\
& = E  \left\{
       1 - \left(
          \frac{1}{3!}  T_{\alpha \beta \gamma} \varepsilon_\alpha \varepsilon_\beta \varepsilon_\gamma 
         + \frac{1}{4!}  F_{\alpha \beta \gamma \lambda} \varepsilon_\alpha \varepsilon_\beta \varepsilon_\gamma \varepsilon_\lambda + ... 
       \right)  \right\} \\
& \qquad \qquad \qquad \left.
+ \frac{1}{2!} \left( \frac{1}{3!} T_{\alpha \beta \gamma}  \varepsilon_\alpha \varepsilon_\beta \varepsilon_\gamma + \frac{1}{4!}  F_{\alpha \beta \gamma \lambda}  \varepsilon_\alpha \varepsilon_\beta \varepsilon_\gamma \varepsilon_\lambda + ... \right) ^2 + ...\right\}  \\
&\sim 1 - E\left(\frac{1}{3!}  T_{\alpha \beta \gamma}  \varepsilon_\alpha \varepsilon_\beta \varepsilon_\gamma + \frac{1}{4!} F_{\alpha \beta \gamma \lambda} \varepsilon_\alpha \varepsilon_\beta \varepsilon_\gamma \varepsilon_\lambda + ...\right) \\
& \qquad \qquad +  \frac{1}{2!} E\left( \left(\frac{1}{3!}\right)^2  T_{\alpha \beta \gamma} T_{\lambda \mu \nu} \varepsilon_\alpha \varepsilon_\beta \varepsilon_\gamma \varepsilon_\lambda \varepsilon_\mu \varepsilon_\nu + ...\right) - ... \\
&\sim 1 
  - \frac{1}{4!} F_{\alpha \beta \gamma \lambda} E\left( \varepsilon_\alpha \varepsilon_\beta \varepsilon_\gamma \varepsilon_\lambda \right) 
  + \frac{1}{2!} \left( \frac{1}{3!}\right)^2  T_{\alpha \beta \gamma} T_{\lambda \mu \nu} E\left(  \varepsilon_\alpha \varepsilon_\beta 
         \varepsilon_\gamma \varepsilon_\lambda \varepsilon_\mu \varepsilon_\nu \right) 
  + ...
\end{align*}

This series diverges, but the terms shown can produce a good approximation to $M$ on their own. A similar, often better approximation is given by a cumulant expansion for $log \ M$ ~\citep{Shun1995, McC1987}, where the first higher-order terms are the same as above, but the product is turned into a sum:

\begin{align*}
log \ M & \sim 
- \ell \left(\hat{\mathbf{y}}\right) + \frac{N}{2} log \left(2 \pi\right) - \frac{1}{2} log \vert \mathbf{H} \vert \nonumber \\
& \qquad - \frac{1}{4!} F_{\alpha \beta \gamma \lambda} E \left(\varepsilon_\alpha \varepsilon_\beta \varepsilon_\gamma \varepsilon_\lambda\right) \nonumber \\
& \qquad + \frac{1}{2} \left( \frac{1}{3!} \right)^2 T_{\alpha \beta \gamma} T_{\lambda \mu \nu} E \left(  \varepsilon_\alpha \varepsilon_\beta \varepsilon_\gamma \varepsilon_\lambda \varepsilon_\mu \varepsilon_\nu\right) \nonumber \\
& \qquad + ... \\
& = - \ell \left(\hat{\mathbf{y}}\right) + \frac{N}{2} log \left(2 \pi\right) - \frac{1}{2} log \vert \mathbf{H} \vert + IV + IIIa + IIIb + ...
\end{align*}

where 

\begin{align*}
IV & = - \frac{1}{24} \cdot 3 F_{\alpha \beta \gamma \lambda} H^{-1}_{\alpha \beta} H^{-1}_{\gamma \lambda} \\
IIIa &= \frac{1}{72} \cdot 9 H^{-1}_{\alpha \beta}T_{\alpha \beta \gamma}H^{-1}_{\gamma \lambda}T_{\lambda \mu \nu}H^{-1}_{\mu \nu} \\
IIIb &= \frac{1}{72} \cdot 6 T_{\alpha \beta \gamma} H^{-1}_{\alpha \lambda} H^{-1}_{\beta \mu} H^{-1}_{\gamma \nu} T_{\lambda \mu \nu}
\end{align*}

We use the cumulant expansion, for $log \ M$, because the derivatives are simpler and because in many cases it is more accurate ~\citep{Shun1995}.

Terms IIIa and IIIb involve the same tensors, but the sums are very different. In term IIIa, as long as we have the near-diagonal entries of $\mathbf{H}^{-1}$, we can immediately convert the third-order tensors into vectors $v_{\gamma} = H^{-1}_{\alpha \beta} T_{\alpha \beta \gamma}$. Then $IIIa = \mathbf{v^T H^{-1} v}$. In section \ref{Sparsity} we explain how to compute the near-diagonal entries in $O(np^3)$ time ~\citep{IEEE}; getting $\mathbf{v^T H^{-1} v}$ is even faster because $\mathbf{H}$ is block-tridiagonal.

Term IIIb is different and much more difficult. Each $\mathbf{H^{-1}}$ connects one mode of the first $\mathbf{T}$ to a mode of the second $\mathbf{T}$, and $\mathbf{H^{-1}}$ is full. This means that entries from one time point of the first $\mathbf{T}$ are multiplied by entries from all other time points in the second $\mathbf{T}$, not just entries from the neighboring time points. This would seem to mean that IIIb requires quadratic time (in $n$) to compute. The appendix  shows how it can be done in linear time by using power series and recurrence relations.

Graphically, the difference can be represented this way:

\begin{tikzpicture}[thick]
\label{eq:IIIa vs IIIb}
  \begin{pgfonlayer}{nodelayer}
  \begin{scope}[shift={(0em, 0)}]
  \node[style=plain] at (-8em, 0) {\textit{IIIa =}};
		\node [style=plain] (T0) at (-4.0em,-1.5em) {
		     $\hspace{0.5em}{\mathbf{T}}\hspace{0.5em}$
		     };
		\node [style=plain] (T1) at (4.0em,-1.5em)  {
		    $\hspace{0.5em}{\mathbf{T}}\hspace{0.5em}$ };
		\node [style=plain] (H0) at (0.0em,-1.5em) {
		    $\mathbf{H^{-1}}$
		    };
		\node [style=plain] (H1) at (-4.0em,1.5em) {
		    $\mathbf{H^{-1}}$};
		\node [style=plain] (H2) at (4.0em,1.5em) {
		    $\mathbf{H^{-1}}$};
		\end{scope}
  \end{pgfonlayer}
  \begin{pgfonlayer}{edgelayer}
  \begin{scope}[shift={(0em, 0)}]
		\draw [style=line, very thick] (5.2em,-1.1em) to [out=20,in=270] (6.1em,0.0em) to [out=90,in=340] (5.2em,1.1em);	
		\draw [style=line, very thick] (2.8em,-1.1em) to [out=160,in=270] (1.9em,0.0em) to [out=90,in=200] (2.8em,1.1em);	
		\draw [style=line, very thick] (-5.2em,-1.1em) to [out=160,in=270] (-6.1em,0.0em) to [out=90,in=200] (-5.2em,1.1em);	
		\draw [style=line, very thick] (-2.8em,-1.1em) to [out=20,in=270] (-1.9em,0.0em) to [out=90,in=340] (-2.8em,1.1em);	
		\draw [style=line, very thick] (T0) to (H0);	
		\draw [style=line, very thick] (T1) to (H0);	
    \end{scope}
  \end{pgfonlayer}
%\end{tikzpicture} 
%\begin{tikzpicture}[thick]
  \begin{pgfonlayer}{nodelayer}
  \begin{scope} [shift = {(16em, 0em)}]
    \node[style=plain] at (-2em, 0) {\textit{IIIb =}};
    \node [style=tensor] (T0) at (0, 0) {$\mathbf{T}$};
    \node [style=tensor] (T1) at (8.0em, 0) {$\mathbf{T}$};
    \node [style=plain] (H0) at (4.0em, 2.0em) {
        $\mathbf{H^{-1}}$};
    \node [style=plain] (H1) at (4.0em, 0) {
        $\mathbf{H^{-1}}$};
    \node [style=plain] (H2) at (4.0em, -2.0em) {
        $\mathbf{H^{-1}}$};
    \end{scope}
  \end{pgfonlayer}
  \begin{pgfonlayer}{edgelayer}
	  \draw [very thick] (T0) to (H0.west);
	  \draw [very thick] (T0) to (H1.west);
	  \draw [very thick] (T0) to (H2.west);
	  \draw [very thick] (T1) to (H0.east);
	  \draw [very thick] (T1) to (H1.east);
	  \draw [very thick] (T1) to (H2.east);
  \end{pgfonlayer}
\end{tikzpicture}
%\end{align*}

%You may want to get Mike to do this. Look, you know the right way to do this, which is to accede to the left-right 

Here each line represents a contraction over two modes: one from each tensor, if there are two, or two from a single tensor. This notation will be helpful later when doing more complicated manipulations on term IIIb.

It is worth emphasizing that all three of these terms are invariant under linear transformations. If we make a change of variables $\mathbf{\tilde{y} = B y}$, 
then

\begin{equation*}
\frac{\partial}{\partial \mathbf{\tilde{y}}} = \mathbf{B}^{-1} \frac{\partial}{\partial \mathbf{y}}
\end{equation*}
and the derivatives tensors (at the critical path) become

\begin{align*}
\tilde{H}_{\alpha \beta}  = \frac{\partial^2 \ell}{ \partial \tilde{y}_{\alpha} \partial \tilde{y}_{\beta} } 
&=  H_{\mu \nu} B^{-1}_{\mu \alpha} B^{-1}_{\nu \beta} 
&  \mathbf{\tilde{H}} &= \mathbf{B^{-T}} \mathbf{H} \mathbf{B}^{-1} \\
\tilde{T}_{\alpha \beta \gamma} 
&= T_{\mu \nu \rho} B^{-1}_{\mu \alpha} B^{-1}_{\nu \beta} B^{-1}_{\rho \gamma} 
& \mathbf{\tilde{T}} &= (\mathbf{B^{-T}}  \otimes \mathbf{B^{-T}}   \otimes \mathbf{B^{-T}})(\mathbf{T}) \\
\tilde{F}_{\alpha \beta \gamma \lambda} 
&= F_{\mu \nu \rho \sigma} B^{-1}_{\mu \alpha} B^{-1}_{\nu \beta} B^{-1}_{\rho \gamma}  B^{-1}_{\sigma \lambda}
& \mathbf{\tilde{F}} &= (\mathbf{B^{-T}}  \otimes \mathbf{B^{-T}}   \otimes \mathbf{B^{-T}} \otimes \mathbf{B^{-T}} ) (\mathbf{F})
\end{align*}
Then the sum 
$- \frac{1}{24}\tilde{F}_{\alpha \beta \gamma \lambda} \tilde{H}^{-1}_{\alpha \beta} \tilde{H}^{-1}_{\gamma \lambda}  $ is still equal to term IV, 
$\frac{9}{72}\tilde{H}^{-1}_{\alpha \beta} \tilde{T}_{\alpha \beta \gamma} \tilde{H}^{-1}_{\gamma \lambda} \tilde{T}_{\lambda \mu \nu} \tilde{H}^{-1}_{\mu \nu}$ is still term IIIa, etc.
In the simplified matrix and graphical notation, the transformation of $\mathbf{T}$ and $\mathbf{F}$ can be written as

\begin{equation*}
\begin{tikzpicture}[thick]
  \begin{pgfonlayer}{nodelayer}
%	  \node [style=plain] (E0) at (-10.0em,0) {$\hat{H}=D^{-1}H D^{T}$};
    \node [style=plain] (That) at (-3.0em, 0) {$\tilde{\mathbf{T}}$\hspace{0.6em}$=$};
    \node [style=plain] (T1) at (2.5em, 0em) {$\mathbf{T}$};
		\node [style=plain] (D00) at (-0.5em, 0em) {$\mathbf{B^{-T}}$};
		\node [style=plain] (D01) at (4.5em, 2.0em) {$\mathbf{B^{-1}}$};
		\node [style=plain] (D02) at (4.5em, -2.0em) {$\mathbf{B^{-1}}$};
		\node [style=plain] (Fhat) at (8.0em, 0em) {$\tilde{\mathbf{F}}$
		             \hspace{0.6em}$=$};
    \node [style=plain] (F1) at (12.0em, 0em) {$\mathbf{F}$};
		\node [style=plain] (D10) at (10.5em, 2.0em) {$\mathbf{B^{-T}}$};
		\node [style=plain] (D11) at (13.5em, 2.0em) {$\mathbf{B^{-1}}$};
		\node [style=plain] (D12) at (10.5em, -2.0em) {$\mathbf{B^{-T}}$};
		\node [style=plain] (D13) at (13.5em, -2.0em) {$\mathbf{B^{-1}}$};
  \end{pgfonlayer}
  \begin{pgfonlayer}{edgelayer}
	  \draw [very thick] (T1) to (D00.east);	
	  \draw [very thick] (T1) to (D01.south west);
	  \draw [very thick] (T1) to (D02.north west);
	  \draw [very thick] (F1) to (D10.south);
	  \draw [very thick] (F1) to (D11.south);
	  \draw [very thick] (F1) to (D12.north);
	  \draw [very thick] (F1) to (D13.north);
  \end{pgfonlayer}
\end{tikzpicture}
\label{eq:threeFormulas}
\end{equation*}

In reality, both tensors are being multiplied symmetrically in all modes. The reason for using transposes is because if a matrix is on the left side, its rows are multiplied, and if a matrix is on the right side, its columns are multiplied. This is consistent with the usual direction of matrix multiplication. Using the same convention will make things simpler in the Appendix.

%\begin{align*}
%\mathbf{\tilde{H}} &= \mathbf{B}^{-T} \mathbf{H} \mathbf{B}^{-1} \\
%\mathbf{\tilde{T}} &= (\mathbf{B}^{-T}  \otimes \mathbf{B}^{-T}   \otimes \mathbf{B}^{-T})(\mathbf{T}) %  \mathbf{H} \mathbf{B}^{-1})
%\\
%\mathbf{\tilde{F}} &= (\mathbf{B}^{-T}  \otimes \mathbf{B}^{-T}   \otimes \mathbf{B}^{-T} \otimes \mathbf{B}^{-T} ) (\mathbf{F})
%\end{align*}

\section{The Sparsity Structure and How to Use It}
\label{Sparsity}
Take another look at the log likelihood $\ell (\mathbf{y})$ (\ref{ell}). Each term in $\ell_{data}$ uses values from one time point. Each term in $\ell_{dyn}$ uses values from two adjacent time points.

This means that if $i$ is any time point, the blocks $\mathbf{H_{i,i}}$, $\mathbf{H_{i , i+1}}$, and $\mathbf{H_{i,i-1}}$ can be nonzero. For any other $j$, i.e. if  $ \vert i-j \vert > 1$, $\mathbf{H_{i,j}}  = \frac{\partial^2 \ell}{\partial \mathbf{y}_i \partial \mathbf{y}_j} = \mathbf{0_{p \times p}}$. Such a matrix is called block-tridiagonal (see figure ~\ref{fig:BOD}).

Likewise, $\mathbf{T_{i,j,k} }$ is zero unless $\max(|i-j|, |j-k|, |i-k|) \le 1$. And $\mathbf{F_{i, j, k, l}}$ is zero if any of the $\binom{4}{2} = 6$ differences is greater than 1.

\subsection{Setup, block-diagonals, block off-diagonals, levels}

In this subsection we see how to work with $\mathbf{H}$, then apply that to computing $IV$ and $IIIa$. The details for $IIIb$ are worked out in the appendix.

%DONE WITH BASIC SPARSITY EXPLANATIONS, ON TO COMPUTATION

If $\hat{\mathbf{y}}$ is a true local minimum of $\ell$, then $\mathbf{H}$ has to be positive definite. This is crucial, because we need the Cholesky decomposition for the calculation.

\begin{figure*}
%\label{fig:Cholesky}
\begin{eqnarray*}
\mathbf{H}  &=&  
\left(
%%%% MATRIX OF H %%%%
\begin{array}{cccc}
\mathbf{P_1} & \mathbf{Q_1}^T & \mathbf{0} & \mathbf{0} \\
\mathbf{Q_1} & \mathbf{P_2} & \mathbf{Q_2}^T & \mathbf{0} \\
\mathbf{0} & \mathbf{Q_2} & \mathbf{P_3} & \mathbf{Q_3}^T \\
\mathbf{0} & \mathbf{0} & \mathbf{Q_3} & \mathbf{P_4} \\
\end{array} \right) \\
&=&  \mathbf{L L^T}, \mathrm{where} \\
\mathbf{L} &=& 
\left(
\begin{array}{cccc}
\mathbf{D_1} & \mathbf{0} & \mathbf{0} & \mathbf{0} \\
\mathbf{E_1} & \mathbf{D_2} & \mathbf{0} & \mathbf{0} \\
\mathbf{0} & \mathbf{E_2} & \mathbf{D_3} & \mathbf{0} \\
\mathbf{0} & \mathbf{0} & \mathbf{E_3} & \mathbf{D_4} \\
\end{array} \right) \\
&=& \mathbf{D + E} \\
&=& \mathbf{D \big(  I + D^{-1} E \big)} \\
&=& \mathbf{D \big(  I - A \big)} \\
\end{eqnarray*}
\caption{Decomposition of $\mathbf{H}$.} \label{fig:Cholesky}
\end{figure*}

Here $\mathbf{D}$ is block-diagonal but not symmetric. In fact, $\mathbf{D}$ is lower-triangular. $\mathbf{A}$ is block off-diagonal, level -1. This means that $\mathbf{A_{i,j}}$ can only be nonzero if $i = j+1$. 

In general, the $np \times np$ matrix $\mathbf{X}$ is block off-diagonal, level $l$, if the blocks $\mathbf{X_{i,j}} \ne \mathbf{0}$ only when $j = i + l$. 

Block-off-diagonals have a raising and lowering property, demonstrated in Figure ~\ref{fig:BOD}. If $\mathbf{X}$ is block off-diagonal with level $l_X$, and $\mathbf{Y}$ is block off-diagonal with level $l_Y$, then $\mathbf{XY}$ is block off-diagonal with level $l_X + l_Y$. ``Raising the level by 1'' happens when you multiply on either size by a block off-diagonal matrix of level 1. ``Lowering by 1'' is the same as ``raising by -1'', which means multiplying by a block off-diagonal of level -1.

\begin{figure*}
%\label{fig:BOD}
If $\mathbf{X}$ has level $l_X=1$, and $\mathbf{Y}$ has level $l_Y=-2$, then $\mathbf{XY}$ has level $ l = l_X + l_Y = -1 $. The third block of level  $\mathbf{XY}$ is zero because $\mathbf{Y}$ has only 2 nonzero blocks. \\

\begin{eqnarray*}
\mathbf{X} \mathbf{Y} &=& 
\left(
\begin{array}{cccc}
\mathbf{0} & \mathbf{X_{1}} & \mathbf{0} & \mathbf{0} \\
\mathbf{0} & \mathbf{0} & \mathbf{X_{2}} & \mathbf{0} \\
\mathbf{0} & \mathbf{0} & \mathbf{0} & \mathbf{X_{3}} \\
\mathbf{0} & \mathbf{0} & \mathbf{0} & \mathbf{0} \\
\end{array} \right) 
\left(
\begin{array}{cccc}
\mathbf{0} & \mathbf{0} & \mathbf{0} & \mathbf{0} \\
\mathbf{0} & \mathbf{0} & \mathbf{0} & \mathbf{0} \\
\mathbf{Y_{1}} & \mathbf{0} & \mathbf{0} & \mathbf{0} \\
\mathbf{0} & \mathbf{Y_{2}} & \mathbf{0} & \mathbf{0} \\
\end{array} \right) \\
{} &=&
\left(
\begin{array}{cccc}
\mathbf{0} & \mathbf{0} & \mathbf{0} & \mathbf{0} \\
\mathbf{X_{2}Y_{1}} & \mathbf{0} & \mathbf{0} & \mathbf{0} \\
\mathbf{0} & \mathbf{X_{3}Y_{2}} & \mathbf{0} & \mathbf{0} \\
\mathbf{0} & \mathbf{0} & \mathbf{0} & \mathbf{0} \\
\end{array} \right)
\end{eqnarray*}
\caption{Off-diagonal matrix multiplication} \label{fig:BOD}
\end{figure*}

\subsection{Computing near-diagonal elements of $\mathbf{H^{-1}}$}
\label{near-diagonals}

%To simplify this, we make our first change of coordinates:

Moving on with the computation,

\begin{eqnarray*}
\mathbf{H} &=& \mathbf{D \big(I - A \big) \big(  I - \mathbf{A} \big)^T D^T  } \\
\mathbf{H^{-1}} &=& \mathbf{D^{-T} \big(I - \mathbf{A} \big)^{-T} \big(  I - A \big)^{-1} D^{-1} } \\
\end{eqnarray*}

The power series for $\mathbf{(I-A)}^{-1}$ actually terminates because $\mathbf{A}$ is strictly lower-triangular. So we can expand and group terms:
%, i.e. $\mathbf{A}^N=0$ (in fact $\mathbf{A}^n=0$ because of the blocking). 

\begin{eqnarray*}
\mathbf{H} &=& \mathbf{D^{-T} }\Big( \sum_{q \ge 0} \mathbf{A^{Tq}} \Big)  \Big( \sum_{r \ge 0} \mathbf{A^{r}} \Big) \mathbf{D^{-1}} \\
&=&  \mathbf{D^{-T}} \Big( \sum_{q \ge 0}  \sum_{r \ge 0} \mathbf{A^{Tq}  A^{r}}  \Big)\mathbf{D^{-1}}
\end{eqnarray*}

The terms of the double sum can be grouped by level so as to give us the near-diagonal levels of $\mathbf{H^{-1}}$. This results in an algorithm similar to the block-tridiagonal case of ~\citep{IEEE}.

Each term in the double sum has exactly one level, $q-r$. Now we will group the terms by level. Break the sum into the two cases, $q < r$ and $q \ge r$:

\begin{align*}
\sum_{q \ge 0}  \sum_{r \ge 0} \mathbf{A^{Tq}  A^r} &= \sum_{r > q \ge 0} \mathbf{A^{Tq}  A^r} + \sum_{q \ge r \ge 0} \mathbf{A^{Tq}  A^r}  \nonumber \\
&= \sum_{\mu > 0, q \ge 0} \mathbf{A^{Tq} A^{q+\boldsymbol{\mu}} }+ \sum_{\nu \ge 0, r \ge 0} \mathbf{ A^{T(r+ \boldsymbol{\nu} )} A^r}  \nonumber \\
&= \sum_{\mu > 0} \Big(\sum_{ q \ge 0} \mathbf{A^{Tq}}  \mathbf{A}^q \mathbf{A ^{\boldsymbol{\mu}}} \Big) + \sum_{\nu > 0} \Big(\sum_{ r \ge 0}  \mathbf{A^{T \boldsymbol{\nu}}} \mathbf{A^{Tr}}   \mathbf{A^{r}}  \Big)  \\
&= \sum_{\mu > 0} \mathbf{S} \mathbf{A ^ {\boldsymbol{\mu}}}  + \mathbf{S} +  
\sum_{\nu > 0} \mathbf{A^{T \boldsymbol{\nu}}}  \mathbf{S} 
\end{align*}
where

\begin{equation*}
\mathbf{S} = \sum_{q \ge 0} \mathbf{A^{Tq} A^q}. 
\end{equation*}

%Note that every term of $\mathbf{S}$ is on the main block-diagonal (level 0). That means that $\mathbf{S}$ is block-diagonal. 
%Moreover, each term in the above sum for 
%$\mathbf{ \big(I - A \big)^{-1} \big(  I - \mathbf{A} \big)^{-T} }$ 
%occupies a single level: $\mathbf{S}$ is level 0, 
%$\mathbf{S A^} \boldsymbol{\mu}$ is level $-\mu$, and 
%$\mathbf{A}^{T \boldsymbol{\nu}} \mathbf{S}$ 
%is level $\nu$.

Now we will show how to compute $\mathbf{S}$ in $O(n p^3)$ time by a block matrix recurrence relation.

Obviously,

\begin{equation*}
\mathbf{S} = \mathbf{I} + \mathbf{A^T SA} \\
\end{equation*}

The map $S \rightarrow \mathbf{A^T SA}$ has a special property: each block of $\mathbf{A^T SA}$ only depends on the \textit{following} block of $\mathbf{S}$, and the last block of $\mathbf{A^T SA}$ is zero.

\begin{align*}
\left(
\begin{array}{ccccc}
 \mathbf{0} & \mathbf{A_1}^T & \mathbf{\cdots} & \mathbf{0} & \mathbf{0} \\
\mathbf{0} & \mathbf{0} & \mathbf{A_2}^T & \mathbf{0} & \mathbf{0} \\
\mathbf{\vdots} & \mathbf{\vdots} & \mathbf{\ddots} & \mathbf{\ddots} & \mathbf{\vdots} \\
\mathbf{0} & \mathbf{0} &  \mathbf{\cdots} & \mathbf{0} & \mathbf{A_{n-1}^T} \\
\mathbf{0} & \mathbf{0} & \mathbf{\cdots} & \mathbf{0} & \mathbf{0}  \\
\end{array} \right) \cdot
%%%%%%%%
%Second matrix
%%%%%%%%
\left(
\begin{array}{ccccc}
 \mathbf{S_1} & \mathbf{0} & \mathbf{\cdots} & \mathbf{0} & \mathbf{0} \\
\mathbf{0} & \mathbf{S_2} & \mathbf{\cdots} & \mathbf{0} & \mathbf{0} \\
\mathbf{\vdots} & \mathbf{\vdots} & \mathbf{\ddots} & \mathbf{\vdots} & \mathbf{\vdots} \\
\mathbf{0} & \mathbf{0} &  \mathbf{\cdots} & \mathbf{S_{n-1}} & \mathbf{0} \\
\mathbf{0} & \mathbf{0} & \mathbf{\cdots} & \mathbf{0} & \mathbf{S_n} \\
\end{array} \right) \cdot
%%%%%%%
%Third matrix
%%%%%%%
\left(
\begin{array}{ccccc}
\mathbf{0} & \mathbf{0} & \mathbf{\cdots} & \mathbf{0} & \mathbf{0} \\
\mathbf{A_1} & \mathbf{0} & \mathbf{\cdots} & \mathbf{0} & \mathbf{0} \\
\mathbf{\vdots} & \mathbf{A_2} & \mathbf{\ddots} & \mathbf{\vdots} & \mathbf{\vdots} \\
\mathbf{0} & \mathbf{0} &  \mathbf{\ddots} & \mathbf{0} &  \mathbf{0}\\
\mathbf{0} & \mathbf{0} & \mathbf{\cdots} & \mathbf{A_{n-1}} & \mathbf{0}  \\
\end{array} \right) 
\\
%End of third matrix
%Start of new line
=\left(
\begin{array}{ccccc}
 \mathbf{A_1^T S_2 A_1} & \mathbf{0} & \mathbf{\cdots} & \mathbf{0} & \mathbf{0} \\
\mathbf{0} & \mathbf{A_2^T S_3 A_2} & \mathbf{\cdots} & \mathbf{0} & \mathbf{0} \\
\mathbf{\vdots} & \mathbf{\vdots} & \mathbf{\ddots} & \mathbf{\vdots} & \mathbf{\vdots} \\
\mathbf{0} & \mathbf{0} &  \mathbf{\cdots} & \mathbf{A_{n-1}^T S_n A_{n-1}} & \mathbf{0} \\
\mathbf{0} & \mathbf{0} & \mathbf{\cdots} & \mathbf{0} & \mathbf{0} \\
\end{array} \right) 
\end{align*}

Because of this property, we can solve $\mathbf{S} = \mathbf{I + A^T SA}$ by a reverse iteration:

\begin{align*}
\mathbf{S_n} &= \mathbf{I} \\
\mathbf{S_k} &= \mathbf{I} + \mathbf{A_{k}^T S_{k+1} A_{k} } \\
\end{align*} for $k$ going from $n-1$ down to 1.

This gives a total of $O(n)$ block multiplications, each of which takes up to $O(p^3)$ flops for total complexity $O(n p^3)$. There may be special cases when there are less than $O(p^3)$ flops per block, but that would be somewhat unusual because $\mathbf{A}$ is generated by a Cholesky decomposition.

Having computed $\mathbf{S}$, we now have

\begin{equation*}
\label{Hinv_sum}
\mathbf{ H^{-1} } = \mathbf{ D^{-T} } \Big(  \sum_{\mu > 0} \mathbf{S} \mathbf{A ^{\boldsymbol{\mu}}}  + \mathbf{S} +  \sum_{\nu > 0} \mathbf{ A ^ {T \boldsymbol{\nu}}S} \Big)  \mathbf{ D^{-1} }
\end{equation*}

We will never actually compute all the entries of $\mathbf{ H^{-1} } $ -- it is full and would need too many flops. For terms IV and IIIa, we only need the block-tridiagonal part,

\begin{equation*}
\mathbf{ D^{-T} } \Big(  \mathbf{S} \mathbf{A}  + \mathbf{S} +  \mathbf{A^T S} \Big)  \mathbf{ D^{-1} },
\end{equation*}
which has $O(np^2)$ entries and takes $O(np^3)$ time to compute.

\subsection{Computing IV and IIIa}

Term IV is simple: each nonzero element of $\mathbf{F}$ occurs exactly once, and is multiplied by block-tridiagonal elements of $\mathbf{H}^{-1}$. $\mathbf{F}$ has $O(np^4)$ nonzero terms, and the number of operations is clearly bounded by $O(np^4)$. It may be less if the blocks of $\mathbf{F}$ are themselves sparse.

Term IIIa is slightly more complicated. The first part $H^{-1}_{\alpha \beta} T_{\alpha \beta \gamma}$ is comparable to term IV: each nonzero element of $\mathbf{T}$ occurs once. This takes up to $O(np^3)$ time and returns a vector $ \mathbf{v}$ with $v_{\gamma} = H^{-1}_{\alpha \beta} T_{\alpha \beta \gamma}$. 
Then term IIIa is proportional to ${v_\gamma} H^{-1}_{\gamma \lambda} v_{\lambda} = \mathbf{v}^T  \mathbf{L}^{-T}  \mathbf{L}^{-1}  \mathbf{v} = \vert \vert  \mathbf{L}^{-1}  \mathbf{v} \vert \vert^2$. Solving $ \mathbf{L}^{-1}  \mathbf{v}$ takes $O(np^2)$ time since $ \mathbf{L}$ is triangular and block-banded.

\section {Numerical Examples and Experiments}

\subsection{An Example: Poisson-Distributed Growth and the Need for Variance Stabilization}
\label{sec:bacteria}
Imagine a bacteria colony with a known rate of division per minute, $\theta$. The deterministic equation of growth would be

\begin{equation*}
\frac{dy}{dt} = \theta y
\end{equation*}

But we know that the process is not really deterministic, and that the bacteria count and number of divisions are integers. So a reasonable model is that the number of divisions over the short time $\Delta_i t$ is Poisson distributed with mean $\theta y_i \Delta_i t$:

\begin{align*}
\Delta_i y &\sim Poiss(\theta y_i \Delta_i t) \\
E (\Delta_i y) &= Var(\Delta_i y) = \theta y_i \Delta_i t
\end{align*}

But SLAM can't handle discrete variables directly: the Laplace approximation uses an integral over the values of the $y_i$. The obvious fix is to use the normal approximation to the Poisson distribution:

\begin{align*}
\Delta_i y &\sim \mathcal{N} (\theta y_i \Delta_i t, \theta y_i \Delta_i t) \\
\ell_{trans} (y_{i+1} | y_i) &= \frac{1}{2} \Big(  log(2 \pi \theta y_i \Delta_i t) + \frac{(\Delta_i y - \theta y_i \Delta_i t)^2}{\theta y_i \Delta_i t} \Big) \\
\ell (\mathbf{y}) &= \sum_{i \in I_{data}} \ell_{data} (y_i | y_i^\ast)+ \frac{1}{2} \sum_{i < N} log(2 \pi \theta y_i \Delta_i t )+ \frac{1}{2} \sum_{i < N} \frac{(\Delta_i y - \theta y_i \Delta_i t)^2}{\theta y_i \Delta_i t}  \\
\end{align*}

where $e^{-\ell_{data}(y_i | y_i^\ast)}$ is the data-based likelihood of $y_i$ given $y_i^\ast$. For now we assume that it peaks at $y_i^\ast$.

Now we should be able to find the critical path $\mathbf{\hat{y}}$ that minimizes $\ell(\mathbf{y})$, take derivatives, and compute the various Laplace terms. And presumably, for small enough time steps, this would give a good approximation to the stochastic differential equation.

\begin{figure}
\includegraphics[scale=0.4]{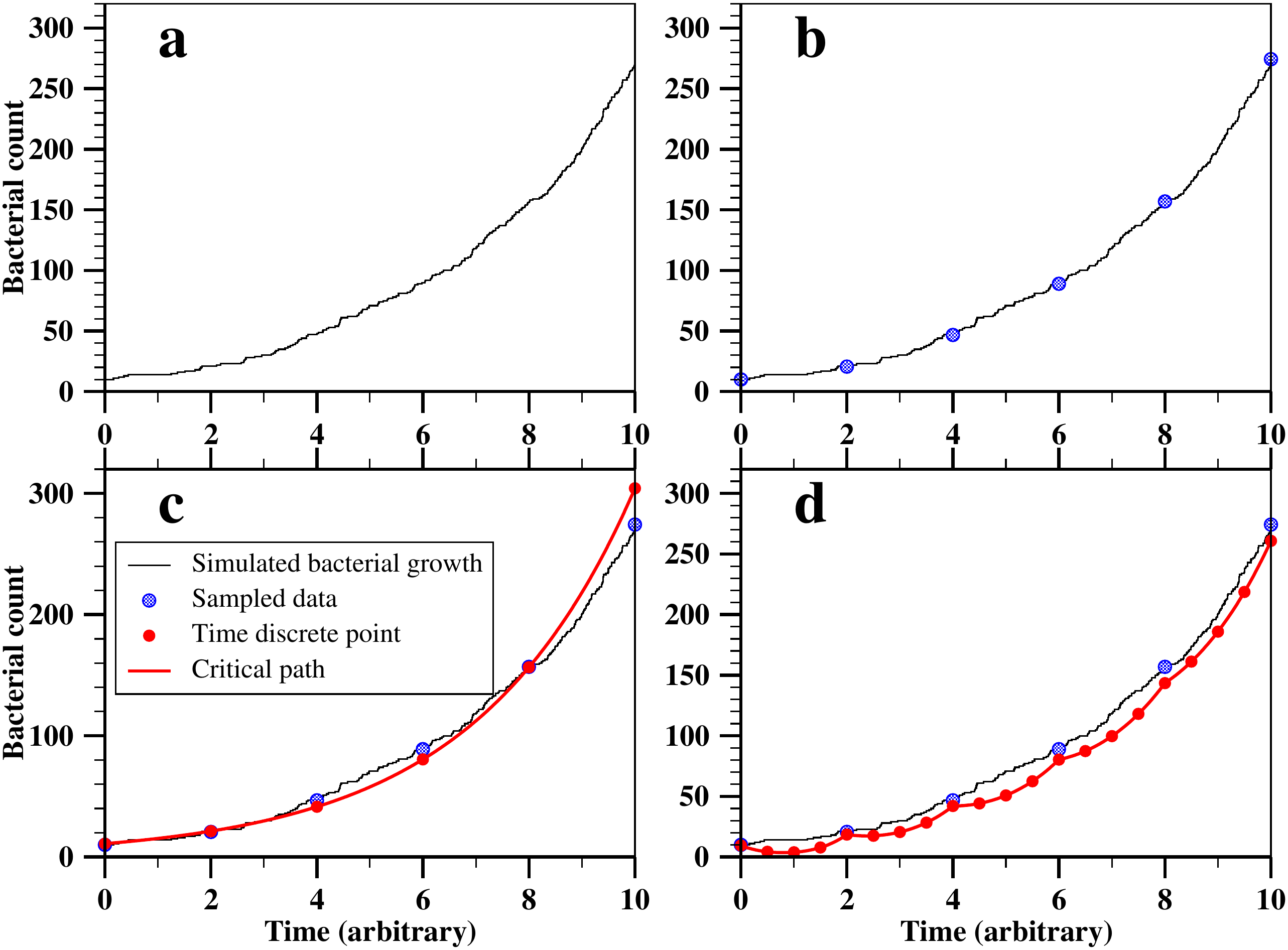}
\caption{Poisson bacterial count figure.}\label{fig:bacteria}
\end{figure}

Unfortunately, without a further tweak, even this simple example can fail precisely \textit{because} of using small time steps between the data points.  If there are too many consecutive time points between the data times, i.e. too many points without data, you can get the problem shown in figure \ref{fig:bacteria}. Increasing $N$ tends to drive $\hat{\mathbf{y}}$ lower and lower; if there are enough points between data points, $\hat{\mathbf{y}}$ can approach zero for some time points.  

This doesn't just mean that the critical paths get ugly; they are actually getting farther and farther from any realistic paths that the system might take. This is possible because the probability density of the path is large for $y$ close to zero, even though the total probability measure of that region is still quite small. The problem stops being good for a Laplace-type approach. 

Fortunately, there is a way around this problem. We start by identifying the reason for it, then explain the solution for this example. 

The explanation below is a heuristic, not a theorem, but it is based on actual observation. Rearranging the expression for $\ell(\mathbf{y})$, we get

\begin{align*}
\ell(\mathbf{y}) &= \frac{1}{2} \sum_{i < N} log(2 \pi \Delta_i t )  + \sum_{i \in I_{data}} \ell_{data} (y_i) +  \frac{1}{2} \sum_{i < N} log(y_i) + \frac{1}{2} \sum_{i < N} \frac{(\Delta_i y - \theta y_i \Delta_i t)^2}{\theta y_i \Delta_i t}  \\
\end{align*}

%To think about this for increasing $N$, we need a sequence of partitions and critical paths. Let $P^{(N)}= \{ t_1^{(N)}, t_2^{(N)}, ..., t_N^{(N)}\}$ be the $N^{th}$ partition of the time interval (including all of the time points with data!) and let $\mathbf{y}^{(N)}$ be the critical path for $P^{(N)}$. Lastly, assume that for all $N$, the numerical derivatives are uniformly bounded, as they would be if there were a continuously differentiable limiting path:
%
%\begin{equation*}
%\frac{\Delta_i y^{(N)}}{\Delta_i t^{(N)}} < M \quad \forall i, N
%\end{equation*}
%

The first term is constant with respect to $\mathbf{y}$ and does not affect the minimization of $\ell({\mathbf{y}})$. The next term is a data term: it only depends on $y_i$ at the data points, and it actually increases when $y_i$ goes below from $y_i^\ast$. So to understand the evolution of the critical path with increasing $N$, we have to look to the other two terms.

The third term (the $log(y)$ term) typically increases without bound as $N \rightarrow \infty$. For example, if there is a limiting path $y(t)$, with evenly spaced time points,  $\sum_{i < N} log(y_i) \sim N \langle log(y) \rangle $ where $\langle log(y) \rangle$ is the mean of $log \ y(t)$.

But the last term does not increase without bound. In practice, 

\begin{equation*}
\frac{1}{2} \sum_{i < N} \frac{(\Delta_i y - \theta y_i \Delta_i t)^2}{\theta y_i \Delta_i t}
\end{equation*}

behaves like a Riemann sum approaching its limiting value--an integral. This is because, in practice, for $\Delta_i t$ small,
%$\Delta_i y \sim O(y_i \Delta_i t)$ 
$\Delta_i y - \theta y_i \Delta_i t \sim O(\Delta_i t)$. (If all of the critical paths were along the same smooth function $y(t)$, then $\Delta_i y - \theta y_i \Delta_i t$ would be $O(\Delta_i t^2)$. This doesn't happen here because the $y_i$ come from critical paths for different partitions. Essentially this sum behaves more like it would for a Brownian motion than for a smooth function.) 

Using $\Delta_i y - \theta y_i \Delta_i t \sim O(\Delta_i t)$, we get
\begin{align*}
\sum_{i=1}^{N-1} \frac {(\Delta_i y -\theta y_i \Delta_it )^2} {\theta y_i \Delta_i t} 
 & \sim \sum_{i=1}^{N-1} \frac {(O(\Delta_i t)^2)} {\theta y_i \Delta_i t} \\ 
 & = \sum_{i=1}^{N-1} O(\Delta_i t)
\end{align*}

if the $y_i$ are uniformly bounded below (uniformly for different $N$).

For this reason, as $N \rightarrow \infty$, the log-likelihood is dominated by the third term, the sum over values of $log \ y_i$. This is why the critical path keeps being pushed lower and lower in order to minimize $\ell$.

This creates a problem: for accuracy with a continuous process, we may need small time steps, but with small time steps we get this artifact. It arises because the variance of $y_{i+1}$ depends on $y_i$: a smaller $y_i$ gives a smaller variance, which gives a greater likelihood. 

The solution is to make a change of variables in the integral.  If we pick the right transformation $v = v(y)$, we can stabilize the variance of $v$ and get a realistic critical path to expand around. If $v_i = v(y_i)$, and $Var(y_i) = \theta y_i \Delta_i t$ is small,

\begin{align*}
Var(v_{i+1}) & \sim v'(y_i)^2 \cdot Var(y_{i+1}) \\
&= v'(y_i)^2 \cdot y_i \Delta_i t
%Var(\Delta (\sqrt{y})) &= Var \Big( \frac{\Delta y}{2 \sqrt{y}} + O(\Delta t^2) \Big) \\
% &= \frac{1}{4y} Var(\Delta y) + O(\Delta t^2) \\
% & \sim \frac{1}{4y} \theta y \Delta t \\
% &= \frac{1}{4} \theta \Delta t \\
\end{align*}
so we can stabilize the variance of $v$ if

\begin{align*}
v'(y)^2 y &= 1 \\
v'(y)^2 &= \frac{1}{y} \\
v'(y) &= \frac{1}{\sqrt{y}} \\
v &= 2 \sqrt{y}
\end{align*}
So $v_i = 2 \sqrt{y_i}$ has stable variance. 
How does this affect the integral for $M(\boldsymbol{\theta})$ ?

If we let 

\begin{equation*}
F(\mathbf{y}) = \frac{1}{2} \sum_i \frac{(\Delta_i y - \theta y_i \Delta_i t)^2}{\theta y_i \Delta_i t}
\end{equation*}
then 

\begin{equation*}
L_{dyn}(\mathbf{y}) \propto \Big( \prod_i \frac{1}{\sqrt{y_i}} \Big) e^{-F(\mathbf{y})} = e^{-F(\mathbf{y})} \Big( \prod_i \frac{1}{\sqrt{y_i}} \Big) 
\end{equation*}

and the marginalized likelihood is proportional to

\begin{align*}
\int L_{data} (\mathbf{y})   e^{-F(\mathbf{y})} \Big( \prod_{i<N} \frac{1}{\sqrt{y_i}} \Big) d^N \mathbf{y}
   &= \int L_{data} (\mathbf{y})  e^{-F(\mathbf{y})} \prod_{i<N} \frac{dy_i}{\sqrt{y_i}} \cdot dy_N\\
&= \int L_{data} (\mathbf{y}) e^{-F(\mathbf{y})} \sqrt{y_N} \prod_{i \le N} d \ (2 \sqrt{y_i})  \\
&= \int L_{data} \Big(\mathbf{y(v)} \Big) e^{-F(\mathbf{y(v)})} \cdot \frac{v_N}{2} \ d^N \mathbf{v} \\
\end{align*}

$F$ is always positive, so $e^{-F} \le 1$.  $L_{data}$ is maximized when $y_i = y_i^\ast$. There is an additional factor of $v_N$, but realistically it cannot go to infinity as fast as $L_{data}$ and $e^{-F}$ go to zero for large $v_N$.  

That means that this integrand, unlike the pre-transformation integrand, is bounded and has some reasonable critical path in terms of $\mathbf{v}$.

\subsection{SIR model}
\label{sec:SIR_model}
Our primary model for testing comes from infectious disease epidemiology. One of the simplest widely-used models is the SIR model ~\citep{SIR1927}. There are three compartments, Susceptibles, Infected, and Recovered. At each time step, some Susceptibles become Infected, and some Infected become Recovered. Two parameters correspond to infectiousness ($\beta$) and speed of recovery ($\gamma$). The traditional method of estimating these parameters uses the deterministic equations

\begin{align*}
\frac{dS}{dt} &= - \beta S I \nonumber \\
\frac{dI}{dt} &= \beta S I - \gamma I \nonumber \\
\frac{dR}{dt} &= \gamma I \nonumber
\end{align*} 

The assumptions behind this are simple: 

(1) new infections are proportional to contacts between susceptible and infected;

(2) new recoveries are proportional to the number infected.

Assumption (2) is memoryless, which is not very realistic, but is often adequate \citep{Anderson}.

To convert this to a stochastic model, we treat new infections at time $i$ ($\nu_{Ii}$) and new recoveries at time $i$ ($\nu_{Ri}$) as independent Poisson variables with means given by the deterministic model:

\begin{eqnarray*} %{rcl}
\mu_{Ii} &=& E \nu_{Ii} = \beta S_i I_i \Delta_i t  \\
\nu_{Ii}  &\sim&  Pois(\beta S_i I_i \Delta_i t) \\
\mu_{Ri} &=& E \nu_{Ri} = \gamma I_i \Delta_i t  \\
\nu_{Ri}  &\sim& Pois(\gamma I_i \Delta_i t )
\end{eqnarray*}

As with the bacterial colony, we use the normal approximation:

%This was an IEEE eqn array.
\begin{eqnarray*}
\nu_{Ii}  & \ \sim \ & \mathcal{N}(\mu_{Ii}, \mu_{Ii})  \\
\nu_{Ri}  & \ \sim \ & \mathcal{N}(\mu_{Ri}, \mu_{Ri}) \\
\Delta_i S & \ = \ & -\nu_{Ii}  \\
& \sim & \mathcal{N}( - \mu_{Ii} ,  \mu_{Ii} )\\
\Delta_i I & \ = \ & \nu_{Ii} - \nu_{Ri} \\
& \ = \   & -\Delta_i S - \nu_{Ri} \\
& \sim & \mathcal{N}(- \Delta_i S - \mu_{Ri}  , \mu_{Ri} )
\end{eqnarray*}

We assume the measurement error is lognormal with a third parameter, $\sigma$:

\begin{equation*}
L_{data} (\mathbf{S}, \mathbf{I} | \sigma) = \prod_{i \in  I_{data}} 
    \frac{1}{2 \pi \sigma^2 S_i^\ast I_i^\ast} 
    \exp \left(- \frac{1}{2} \frac{log(S_i/S_i^\ast)^2}{\sigma^2} \right)  
    \exp \left(- \frac{1}{2} \frac {log(I_i/I^\ast)^2}{\sigma^2} \right)
\end{equation*}

The dynamical part of the likelihood is 

\begin{align*}
L_{dyn} ( \mathbf{S, I} | \beta, \gamma) &= \prod_{i=1}^{n-1}  
    \frac{1}{ \sqrt{2 \pi \mu_{Ii}} }  
    \exp \left( - \frac{1}{2} \frac {(\Delta_i S + \mu_{Ii} )^2} {\mu_{Ii}} \right)  
    \cdot 
    \frac{1} { \sqrt{2 \pi \mu_{Ri}} }  
    \exp \left( - \frac{1}{2} \frac {(\Delta_i I + \Delta_i S + \mu_{Ri} )^2} {\mu_{Ri}} \right)  \nonumber \\
 &= (2 \pi)^{-N/2} \prod_{i=1}^{n-1} \frac{1}{\sqrt{ \mu_{Ii} \mu_{Ri} }}  
           \exp  \Big( -F(S_i, I_i, S_{i+1}, I_{i+1} \vert \beta, \gamma, \Delta_i t ) \Big) \\
 \text{where} \\
 & F(S_i, I_i, S_{i+1}, I_{i+1}) = \frac{1}{2} \frac {(\Delta_i S + \mu_{Ii} )^2} {\mu_{Ii}} 
 \frac{1}{2} \frac {(\Delta_i I + \Delta_i S + \mu_{Ri} )^2} {\mu_{Ri}} \\
 \text{so} \\
 L_{dyn} &= (2 \pi)^{-N/2} 
    \exp \Big( - \sum_{i}^{n-1} F(S_i, I_i, S_{i+1}, I_{i+1}) \Big) 
    \prod_{i=1}^{n-1} \frac{1}{\sqrt{ \mu_{Ii} \mu_{Ri} }} \\
    &= (2 \pi)^{-N/2} e^{-F_{all}(\mathbf{S}, \mathbf{I})} 
    \prod_{i=1}^{n-1} \frac{1}{ \sqrt{\beta S_i I_i \Delta_i t \cdot \gamma I_i \Delta_i t} } \\
&= \frac{(2 \pi)^{-N/2}}{\sqrt{\beta \gamma}} 
e^{-F_{all}(\mathbf{S}, \mathbf{I})} 
\prod_{i=1}^{n-1} \frac{1}{ \sqrt{ S_i } \cdot I_i \Delta_i t } \\
\end{align*}

As with the bacteria model \S\ref{sec:bacteria}, the critical path often gets artifacts for small time steps and the approximation can be poor. So we rewrite the differential as 

\begin{align*}
  L_{dyn} \ d^n S \ d^n I 
  & \propto  \exp(-F_{all}) \prod_{i=1}^{n-1} \frac{dS_i \ dI_i}{\sqrt{ S_i } I_i \Delta_i t } \cdot dS_n dI_n  \\
  & \propto \exp(-F_{all}) \prod_{i=1}^{n-1} \frac{dS_i}{\sqrt{ S_i }} \frac{dI_i}{I_i } \\
\end{align*}

so we want to pick transformed variables $v_S$ and $v_I$ such that

\begin{equation*}
dv_S \propto  \frac{dS}{\sqrt{ S }} , \ dv_I \propto \frac{dI}{I}
\end{equation*}

If we take

\begin{equation*}
v_S = 2 \sqrt{S}, \ v_I = log(I) \\
\end{equation*}

then our integral avoids the artifacts explained in \S\ref{sec:bacteria}.

\subsection{Comparison with Rstan on British Boarding School data}\label{sec:rstan-comparison}
Stan ~\citep{STAN} is a tool for doing Bayesian statistics using Hybrid Monte Carlo estimation. Among other things, it can estimate Bayesian posteriors for parameters in very general models, which can be specified with an easy-to-use general model language. 

This makes Stan a natural check for the approximation methods in SLAM. The probabilistic model in \S\ref{sec:SIR_model} was defined in Stan as well as SLAM, and used to estimate parameters for a classic data set ~\citep{SIRdata}. The purpose is to verify that the estimates comes out similar, but that SLAM can make the calculations more quickly because of the approximation. The posterior median given by Stan is not identical to the maximum marginalized likelihood estimate given by SLAM. However, for small variances, we expect them to be close.

The results are shown in the table below along with the corresponding results from SLAM.
The test data set comes from an influenza infection at a British boarding school ~\citep{SIRdata}. Initial guesses for $\beta$ and $\gamma$ are set equal to Murray's deterministic estimates. The initial guess for $\sigma$ (the measurement error parameter) is arbitrarily set to 0.1.

%This example was given by Murray, based on data compiled by the British Communicable Disease Surveillance
%Centre (British Medical Journal, March 4 1978, p. 587). The event was a flu epidemic in a boys boarding school in the north
%of England. There were 763 resident boys, including one initial infective. The data (given in graphical form only) is for
%residents confined to bed, and, following Murray, we will assume those to be the infectives. The data for the two-week
%epidemic are given in the list below, constructed by reading values from the graph in the original publication.

\begin{figure*}[ht]
\centering
\includegraphics[width=0.6\textwidth]{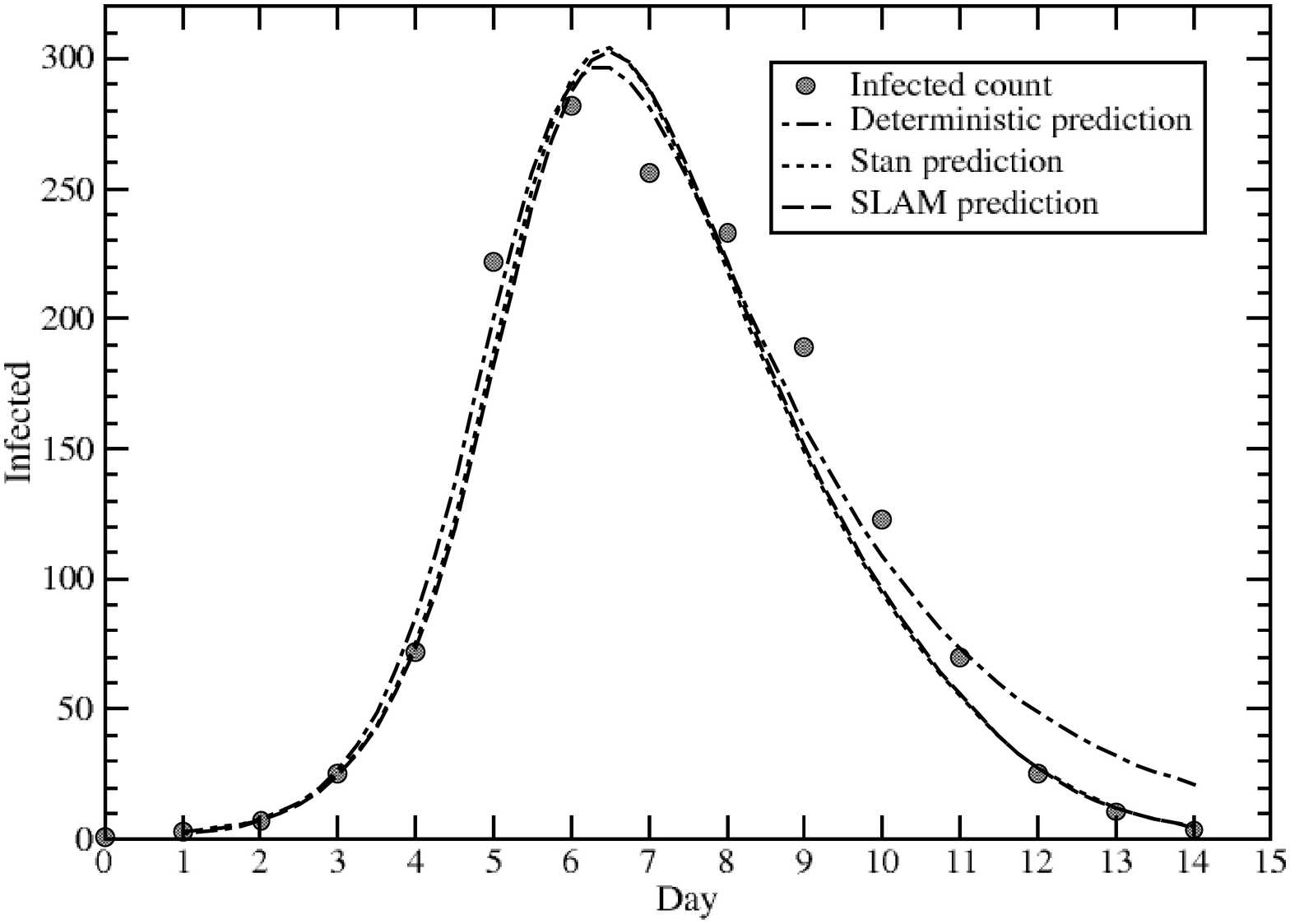}
\caption{Comparison of Stan and SLAM predictions. The deterministic prediction follows the exact ODEs using the best fit values of $\beta$ and $\gamma$ provided by Murray. Stan shows a posterior mean, while SLAM uses a maximum likelihood path.}\label{fig:Stan_vs_SLAM_SIR_I}
\end{figure*}

Figure ~\ref{fig:Stan_vs_SLAM_SIR_I} shows the estimates of the course of infection given by SLAM, Stan, and the deterministic ODEs, given their estimated parameter values. The deterministic model suggests that the epidemic could not end as quickly as it did. Stan and SLAM can follow the data more closely using the stochastic model from \S\ref{sec:SIR_model}. The Stan and SLAM paths are very similar, but we don't expect them  to be identical because Stan is showing posterior means while SLAM is showing a overall maximum likelihood estimate. 

Table ~\ref{tab:StanResults} shows the results of the test. As expected, the parameter estimates from Stan and SLAM are very similar.  For $\beta$ and $\gamma$, the differences are negligible. 
 %within a fraction of a percent of the parameter value; 
For $\sigma$ the Stan and SLAM estimates are within a fraction of the CI, or about 10\% of the value. But for this problem, compared with Stan, basic SLAM was about 35 times faster than Stan, and higher-order SLAM was more than 13 times faster.

\begin{table} 
\begin{tabular}{lcccccc}
{} & \hspace{2mm} & $\beta$ & $\gamma$ & $\sigma$ & \hspace{2mm} & Time\\
\hline
Deterministic (Murray)     & {} & $2.18\times 10^{-3}$ & 0.440 & --   & {} & {} \\
\hline
Stan CI Low       & {} & $2.13\times 10^{-3}$ & 0.462 & 0.092 & {} & {} \\
\hline
Stan posterior median       & {} & $2.48\times 10^{-3}$ & 0.518 & 0.194 & {} & 61.3 $s$ \\
SLAM basic        & {} & $2.47\times 10^{-3}$ & 0.519 & 0.175 & {} & 1.7 $s$ \\
SLAM higher order & {} & $2.47\times 10^{-3}$ & 0.519 & 0.176 & {} & 4.5 $s$ \\
\hline
Stan CI High      & {} & $2.94\times 10^{-3}$ & 0.601 & 0.400 & {} & {} \\
\hline \\
\end{tabular}
\caption{SIR model results.} \label{tab:StanResults}
\end{table}

For this test, Rstan 2.8.0 was run in R 3.2.2, using 4 chains of length 2000 each, with options set to maximize speed (multicore, optimized compilation). Different versions of Stan code for the model were tested. Surprisingly, the fastest version used unvectorized code. 

\subsection{Comparison with INLA on Tokyo rain data}
INLA ~\citep{INLA, INLA-URL} is an R package which also uses Laplace approximations to compute integrals and estimate parameters, but for a different class of problems (Gaussian Markov Random Fields ~\citep{GMRFs}). GMRFs on a line can be analyzed by both INLA and SLAM. 
INLA also uses sparsity, but the higher-order terms are calculated in a simpler (and more general) way which takes quadratic time and memory in the number of nodes. Therefore SLAM, with a linear-time algorithm for the higher-order terms, should be faster for sufficiently large GMRFs on a line. These experiments show that this is true for repeats of the Tokyo data set ~\citep{INLA-MAN} from the INLA package.

The Tokyo data set looks at Tokyo rainfall from the start of 1983 to the end of 1984.  Each row of Tokyo gives a calendar day (e.g. January 28) and the number of times it rained on that day in either 1983 or 1984. For obvious reasons this count is always 2 except for February 29 (from 1984). The top few rows are presented in Table \ref{tab:TokyoRain}.

% latex table generated in R 3.0.2 by xtable 1.7-3 package
% Mon Jan 12 06:44:56 2015
% latex table generated in R 3.0.2 by xtable 1.7-3 package
% Mon Jan 12 07:15:19 2015
\begin{table}[ht]
\centering
\begin{tabular}{ccc}
  \hline
  i & $n_i$ & $y_i$ \\ 
  \hline
1 &   2 &   0 \\ 
 2 &   2 &   0 \\ 
 3 &   2 &   1 \\ 
 4 &   2 &   1 \\ 
 5 &   2 &   0 \\ 
   \hline
\end{tabular}
\caption{Initial rows of Tokyo rain data.} \label{tab:TokyoRain}
\end{table}

Here $i$ is the calendar date (e.g. 1 is January 1), $n_i$ is the number of times that date occurred, and $y_i$ is the number of rain days for that date.

In the INLA manual ~\citep{INLA-MAN}, the Tokyo data set is analyzed using INLA with a second-order random walk model (~\citet{GMRFs}, section 3.4.1). For each date $i$, the chance of rain is assumed to be some $p_i$, so the chance of $y_i$ rain days for that date is $\binom{n_i}{y_i} p_i^{y_i} (1-p_i)^{n_i - y_i}$. The second-order random walk is over $x_i = log(p_i / (1-p_i)$, so $\ell$ for given $\lambda$ is

\begin{equation*}
\ell(\mathbf{x} | \lambda) = \sum_{i=1}^N \Big( - y_i x_i + n_i \ log(1 + e^{x_i})   \Big)
           + \frac{1}{2} \sum_{i=1}^{N-2} \Big( - log \ \lambda +  \frac{\lambda}{\Delta_i t} (x_{i+2} - 2 x_{i+1} + x_i)^2  \Big)
\end{equation*}

After slight modifications\footnote{Three modifications were made in order to work around features that are not yet present in the SLAM code. 

1. SLAM doesn't take second
derivatives, so an extra variable $\hat{x}'_i$ is added which is forced to be nearly equal to the true $x'(t_i)$. Then the usual RW2 penalty is given in terms of $\sum_i (\Delta_i \hat{x}')^2$. 

2. INLA has an option ``cyclic" which forces the function to be periodic. SLAM does not currently have such an option. Therefore cyclic is set to FALSE for INLA. Happily, the predicted functions are nearly periodic.

3. The Tokyo data set covers 1983-84, so $n_i=2$ for all dates except February 29. For Feb. 29 $n_i=1$ because only 1984 was a leap year. For now, it was necessary to set $n_i=2$ for February 29 for both programs. Think of it as a recording error.
}  %end of footnote
we compared the behavior and performance of INLA and SLAM.

In order to adjust the size of the data set, we simply repeated the data set k times, e.g., for k=2 there are 732 data points instead of 366 and $y_{i+366} = y_i$. 

By default, INLA uses an approximation to the higher order terms which is easier to compute (strategy=`simplified.laplace'). To get the full higher order Laplace terms, as used by SLAM, we set INLA strategy = `laplace'.

\begin{figure*}[ht]
\centering
\includegraphics[width=0.6\textwidth]{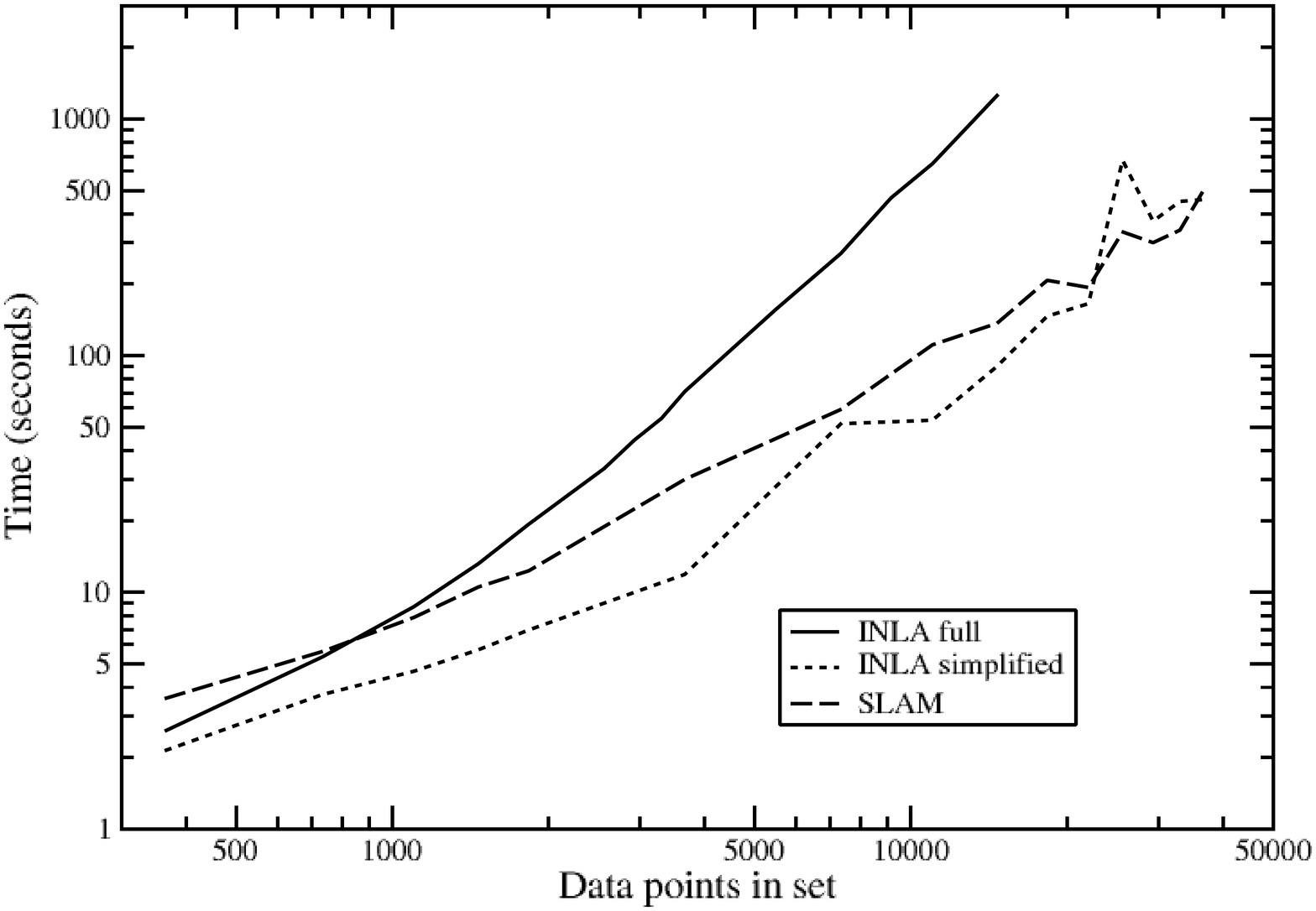}
\caption{Comparison of INLA and SLAM run times.}\label{fig:INLA_SLAM_Log_Times}
\end{figure*}

After running both packages on Tokyo, the estimates are not identical but are very similar, especially for the larger data sets (see figures). The most important results here have to do with run time and especially memory use. For k up to 3, INLA is faster. At k=4 ($N \sim 1.5 \times 10^3$), SLAM becomes faster. The relative difference grows as shown in figure \ref{fig:INLA_SLAM_Log_Times}.

%\begin{table}[ht]
%\centering
%\begin{tabular}{c cccc c}
%  \hline
% & INLA Full & INLA Simplified & INLA Cyclic & SLAM & True \\ 
%  \hline
%Log $\lambda$ & 9.12 & 9.12 & 9.13 & 9.29 & -- \\ 
%  SSE vs Cyclic & $7.9 \times 10^{-4}$ & $3.4 \times 10^{-3}$ & --  & $1.1 \times 10^{-3}$ & -- \\ 
%  Mean p & 0.2586 & 0.2663 & 0.2594 & 0.2623 & 0.2623 \\ 
%   \hline
%\end{tabular}
%\caption{INLA and SLAM estimates using Tokyo rain data.} \label{tab:TokyoRainEstimates}
%\end{table}
%
%
\begin{figure*}[ht]
\centering
\includegraphics[width=0.6\textwidth]{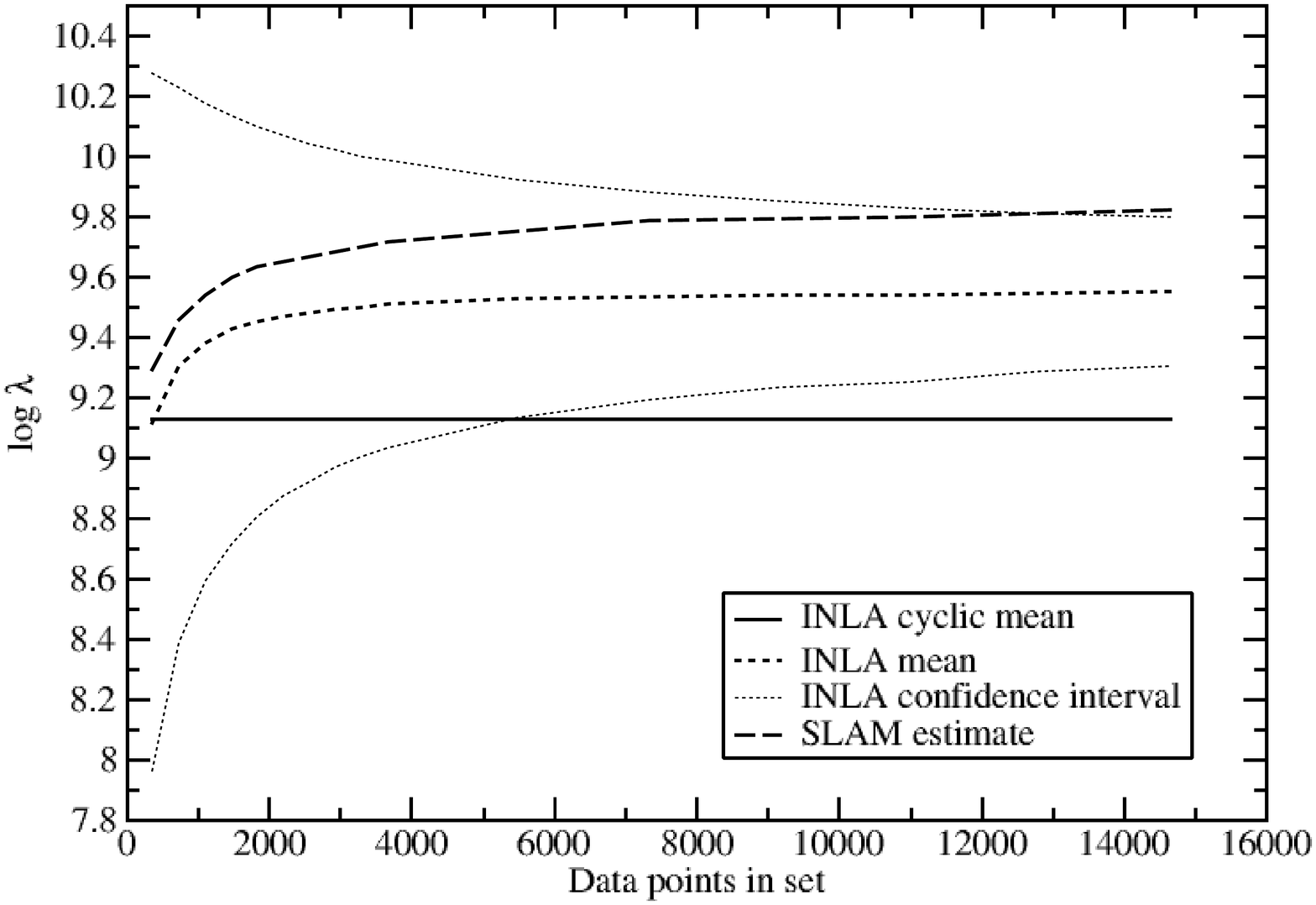}
\caption{Comparison of INLA estimates, the means and ranges.}\label{fig:INLAEstimatesComparisonMeansRanges}
\end{figure*}

\begin{figure*}[ht]
\centering
\includegraphics[width=0.6\textwidth]{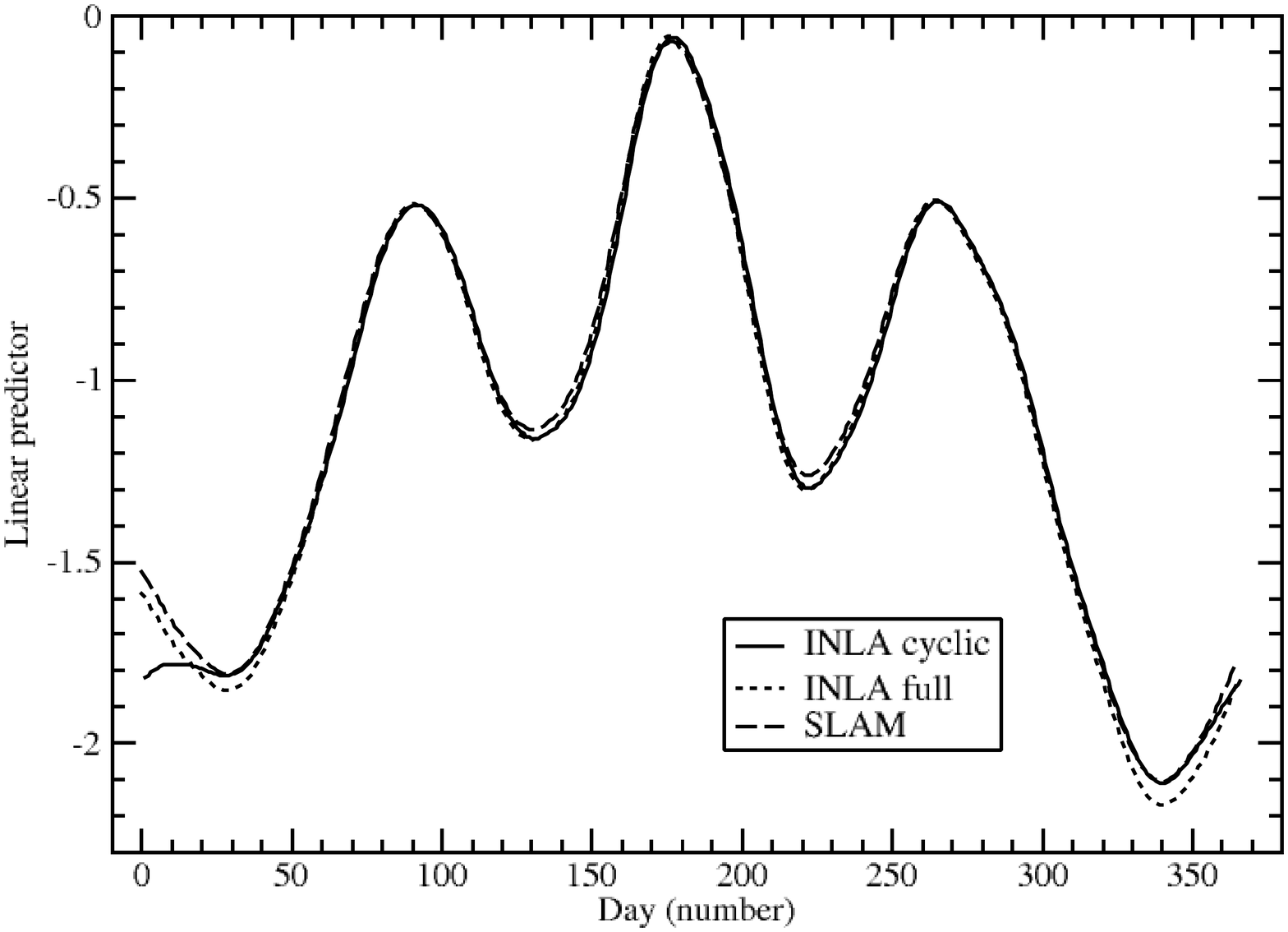}
\caption{Linear predictors.}\label{fig:LinearPredictors1}
\end{figure*}

The SSE for the simplified Laplace is several times larger than for INLA full or for SLAM. This is almost entirely due to greater error near the ends of the year (1 and 366).

The expected number of rain days over the year is
$\sum_i n_i \hat{p_i}$ and the observed number of rain days is $\sum_i y_i$. For SLAM these are equal to at least four digits. This probably means that, under some conditions, the method forces them to be equal. For now we make no suggestion why. 

Lastly, in this example we had to introduce a variable $\hat{x}'$ which is supposed to be the time derivative of the linear predictor $x$. We forced $\hat{x}'$ to be close to the true derivative by adding a large penalty for any difference between $\hat{x}'_i$ and $\Delta_i x / \Delta_i t$. These results show that it is (sometimes) possible to use SLAM for a constrained system using this simple trick.

\section {Discussion}

We have shown how to efficiently compute a higher-order Laplace approximation for functional integrals on time. This can be used to define a marginalized pseudolikelihood for differential equation systems involving randomness. In turn, these approximate marginalized likelihoods can be maximized in order to estimate parameters of the system, such as infectiousness of a disease, or a good roughness parameter for a smoothing problem.
This approximation is implemented in the MATLAB package SLAM. SLAM is compared on real data against two leading packages which also estimate/predict parameters according to a fully Bayesian system or mixed model. As predicted, the time needed for SLAM seems to grow linearly. Compared with either package, for medium-sized data sets using higher-order terms, SLAM produces equivalent results in far less time. This is especially interesting given that both Stan and INLA use compiled, optimized code.
Finally, for some problems, if higher-order terms are used then SLAM is able to process significantly larger data sets than INLA.

\section{Acknowledgements} 
This work was begun at the Center on Aging and Health, Johns Hopkins Bloomberg School of Public Health, under Training Grant T32AG000247 (Epidemiology and Biostatistics of Aging). Additional support was provided by the Biostatistics Department at JHSPH.
I would like to thank Alan Cohen (now at Sherbrooke University) for inviting me into the project which led to this. I would also like to thank Ravi Varadhan and Giles Hooker for consultation and encouragement, and Karen Bandeen-Roche for making it possible. Michael Schumaker provided help with editing and LaTeX. Thanks are also due to the developers of Stan, INLA, and the MATLAB Tensor Toolbox for quick and friendly answers to my questions.

\section {Appendix: Calculation of Term IIIb in linear time}
\label{Appendix}

\subsection{Tensor levels}
This shows how to compute term IIIb in $O(np^4)$ time. This is possible because of the raising and lowering principle described in section \ref{Sparsity}.  A similar concept applies to block tensors. A tensor of order 3, $\mathbf{X}$, is block off-diagonal with level $(l_1, l_2, l_3)$ if every nonzero block $\mathbf{X_{i_1, i_2, i_3}}$ has $i_1 - l_1 =  i_2 - l_2 =  i_3 - l_3$. In other words, $\mathbf{X}$ is level $(l_1, l_2, l_3)$ if it is block off-diagonal and includes the block $(l_1, l_2, l_3)$. Levels are an equivalence class: level $(l_1, l_2, l_3)$ is the same as level $(l_1 + \Delta l, l_2 + \Delta l, l_3 + \Delta l)$ for any integer $\Delta l$. 

The levels of a third-order block tensor are shifted when powers of $\mathbf{A}$ or $\mathbf{A^T}$ are applied to the modes. Our $\mathbf{T}$ is a sum of a handful of block-diagonal and block off-diagonal levels. To compute IIIb, we find the combinations of powers of $\mathbf{A}$ and $\mathbf{A^T}$ which connect the levels of the first $\mathbf{T}$ with the levels of the second $\mathbf{T}$. The grouping is more complicated than for computing the near-diagonals of $\mathbf{H}^{-1}$, but is similar in spirit.

Suppose have a single-level matrix $\mathbf{B}$ and a single-level tensor $\mathbf{T_{mono}}$ with level $(l_1, l_2, l_3)$. If $\mathbf{B}$ is level $l_B$, then the new tensor $(\mathbf{B \otimes I \otimes I}) (\mathbf{T_{mono})}$ is also block off-diagonal, but with level $(l_1 + l_B, l_2, l_3)$. The next observation is critical: if $\mathbf{B}$ is applied to every mode of $\mathbf{T_{mono}}$, the result $\left( \mathbf{B \otimes B \otimes B} \right) (\mathbf{T_{mono}})$ is the same exact level as $\mathbf{T_{mono}}$.

\subsection{A simplifying coordinate transformation}

As we discussed in section ~\ref{Appendix}, we can write term IIIb more simply by making a coordinate transformation:

\begin{equation*}
\begin{tikzpicture}[thick]
  \begin{pgfonlayer}{nodelayer}
    \node [style=tensor] (T0) at (-1.5em, 0) {$\mathbf{T}$};
    \node [style=tensor] (T1) at (9.5em, 0) {$\mathbf{T}$};
    \node [style=tensor] (eq) at (11.5em, 0) {$=$};
    \node [style=plain] (H0) at (4.0em, 2.0em) {$\mathbf{D^{-T}} \tilde{\mathbf{H}}^{-1} \mathbf{D}^{-1}$};
    \node [style=plain] (H1) at (4.0em, 0) {$\mathbf{D^{-T}} \tilde{\mathbf{H}}^{-1} \mathbf{D}^{-1}$};
    \node [style=plain] (H2) at (4.0em, -2.0em) {$\mathbf{D^{-T}} \tilde{\mathbf{H}}^{-1} \mathbf{D}^{-1}$};
  \end{pgfonlayer}
  \begin{pgfonlayer}{edgelayer}
	  \draw [very thick] (T0) to (H0.west);
	  \draw [very thick] (T0) to (H1.west);
	  \draw [very thick] (T0) to (H2.west);
	  \draw [very thick] (T1) to (H0.east);
	  \draw [very thick] (T1) to (H1.east);
	  \draw [very thick] (T1) to (H2.east);
  \end{pgfonlayer}
\end{tikzpicture}
%\label{eq:penrose1}
%\end{equation*}
%\begin{equation*}
%=
\begin{tikzpicture}[thick]
  \begin{pgfonlayer}{nodelayer}
    \node [style=tensor] (T0) at (0, 0.0) {$\tilde{\mathbf{T}}$};
    \node [style=tensor] (T1) at (8.0em, 0.0) {$\tilde{\mathbf{T}}$};
    \node [style=plain] (H0) at (4.0em, 2.0em) {$\tilde{\mathbf{H}}^{-1}$};
    \node [style=plain] (H1) at (4.0em, 0) {$\tilde{\mathbf{H}}^{-1}$};
    \node [style=plain] (H2) at (4.0em, -2.0em) {$\tilde{\mathbf{H}}^{-1}$};
  \end{pgfonlayer}
  \begin{pgfonlayer}{edgelayer}
	  \draw [very thick] (T0) to (H0.west);
	  \draw [very thick] (T0) to (H1.west);
	  \draw [very thick] (T0) to (H2.west);
	  \draw [very thick] (T1) to (H0.east);
	  \draw [very thick] (T1) to (H1.east);
	  \draw [very thick] (T1) to (H2.east);
  \end{pgfonlayer}
\end{tikzpicture}
\label{eq:penrose2}
\end{equation*}
where

\begin{equation*}
\tilde{\mathbf{T}} = (\mathbf{D ^ {-T}} \otimes \mathbf{D ^ {-T}} \otimes \mathbf{D ^ {-T}}) (\mathbf{T}).
\end{equation*}

$\mathbf{D}$ is block-diagonal with block size $p$, so the coordinate transformation takes $O(np^4)$ time.

To continue, we will use the series
\begin{equation*}
\mathbf{ \tilde{H}^{-1} } =  \mathbf{S} \sum_{\mu > 0}  \mathbf{A ^{\boldsymbol{\mu}}}  + \mathbf{S} +  \sum_{\nu > 0} \mathbf{ A ^ {T \boldsymbol{\nu}}S}
\end{equation*}
We simplify this by making a second block-diagonal coordinate transformation to eliminate $\mathbf{S}$. Note that $\mathbf{S}$ is block-diagonal and symmetric because each of its terms $\mathbf{ A^{Tq}  {A}^q}$ is block-diagonal and symmetric. $\mathbf{S}$ is also positive definite, since $\mathbf{I}$ is positive definite and every $\mathbf{ A^{Tq}  {A}^q}$ is semidefinite. So if we define $\mathbf{C}$ to be the (upper!) Cholesky decomposition $\mathbf{S} = \mathbf{C^T C}$.
Then $\mathbf{C^{-T} S} = \mathbf{C}$, 
$\mathbf{S} \mathbf{C}^{-1} = \mathbf{C^T}$ and $\mathbf{C^{-T} S C^{-1}} = \mathbf{I}$. Now we can make the coordinate transformation 

\begin{equation*}
\mathbf{y} \rightarrow \mathbf{C}^{-1} \mathbf{y}, \quad \hat{\mathbf{H}} = \mathbf{C^T \tilde{H} C}
\end{equation*}
from which

\begin{align*}
\mathbf{\hat{H}^{-1}} &= \mathbf{C^{-T} \tilde{\mathbf{H}}^{-1} C^{-1} } \\
&= \mathbf{C^{-T}} \Big( \mathbf{S} \sum_{\mu > 0}  \mathbf{A ^{\boldsymbol{\mu}}}  + \mathbf{S} +  \sum_{\nu > 0} \mathbf{ A ^ {T \boldsymbol{\nu}} S } \Big) \mathbf{C}^{-1} \\
&= \mathbf{C^{-T}}  \Big( \mathbf{S} \sum_{\mu > 0}  \mathbf{A ^{\boldsymbol{\mu}}} \Big) \mathbf{C}^{-1}
      + \mathbf{I} %\mathbf{C^{-T}}  \mathbf{S} \mathbf{C} 
      + \mathbf{C^{-T}} \Big( \sum_{\nu > 0} \mathbf{ A ^ {T \boldsymbol{\nu}}}  \mathbf{S} \Big) \mathbf{C}^{-1} 
%&= \mathbf{C} \sum_{\mu > 0}  \mathbf{A ^{\boldsymbol{\mu}}} \mathbf{C}^{-1}
%      + \mathbf{I} +
%      \mathbf{C^{-T}} \sum_{\nu > 0} \mathbf{ A ^ {T \boldsymbol{\nu}}}  \mathbf{C^T}
\end{align*}

The first and third terms are transposes of each other. Let's look at the first term alone.

\begin{align*}
\mathbf{C^{-T}}  \Big( \mathbf{S}  \sum_{\mu > 0}  \mathbf{A ^{\boldsymbol{\mu}}} \Big) 
\mathbf{C}^{-1}
&= \mathbf{C} \Big( \sum_{\mu > 0}  \mathbf{A ^{\boldsymbol{\mu}}} \Big) \mathbf{C}^{-1} \\
&= \sum_{\mu > 0}  \Big( \mathbf{C}  \mathbf{A ^{\boldsymbol{\mu}}} \mathbf{C}^{-1}  \Big) \\
&= \sum_{\mu > 0}  \hat{\mathbf{A}} ^ {\boldsymbol{\mu}} 
\end{align*}
with $\hat{\mathbf{A}} = \mathbf{C^{-1} A  C}$. At last,

\begin{align*}
\mathbf{\hat{H}^{-1}} &= \sum_{\mu > 0}  \hat{\mathbf{A}} ^ {\boldsymbol{\mu}} + \mathbf{I} + \sum_{\nu > 0}  \mathbf{\hat{A}} ^ {T \boldsymbol{\nu}} \\
&= \sum_{\mu \ge 0}  \hat{\mathbf{A}} ^ {\boldsymbol{\mu}}  + \sum_{\nu \ge 0}  \mathbf{\hat{A}} ^ {T \boldsymbol{\nu}} - \mathbf{I} \\
&= \mathbf{\mathcal{A}} + \mathbf{\mathcal{A}^T} - \mathbf{I} 
\end{align*}
where $\mathbf{\mathcal{A}} = \sum_{\mu \ge 0}  \hat{\mathbf{A}}$.

\subsection{Statement of Results}

\begin{align*}
IIIb 
  & = 2 \langle \hat{\mathbf{T}},  \mathcal{S} (\hat{\mathbf{T}}) \rangle \\
  & \qquad 
  + 6 \langle \mathbf{ 
                    \hat{\mathbf{T}},  
	            \Big( I \otimes (I + \hat{\mathbf{A}} + \mathbf{ \hat{A}^2 }) \otimes     
	                 (\hat{\mathbf{A}} + \mathbf{ \hat{A}^2 }) \Big) (\mathcal{S} (\hat{\mathbf{T}}))
	} \rangle \\
  & \qquad 
  + 6 \langle \mathbf{ 
                     \hat{\mathbf{T}},  
	            \Big( (I + \hat{\mathbf{A}}) \otimes  \hat{\mathbf{A}} \otimes     
	                 \hat{\mathbf{A}}^T \Big) (\hat{\mathbf{T}})
	}\rangle \\
  & \qquad
  - \langle \mathbf{ \hat{\mathbf{T}},  \hat{\mathbf{T}} }\rangle
\end{align*}

where $\langle  \mathbf{P, Q}  \rangle = P_{\alpha \beta \gamma}  Q_{\alpha \beta \gamma}$

and $\mathcal{S} (\hat{\mathbf{T}})$ is defined as 

\begin{equation*}
\mathcal{S} (\hat{\mathbf{T}}) =  \sum_{q \ge 0} \Big( \mathbf{ \hat{A} ^ q} \otimes \mathbf{ \hat{A} ^ q} \otimes \mathbf{ \hat{A} ^ q} \Big) (\hat{\mathbf{T}}).
\end {equation*}

$\mathcal{S} (\hat{\mathbf{T}})$ can be computed in $O(np^4)$ time, and so can the whole sum.

\subsection{How to expand the sum and group the terms}

%Show basic expansion (using regular tensor notation?)
%Work out the coordinate transformations 
%Show how H^{-1} simplifies into the new coordinates
%Define my little "bracket" notation

Let's start by defining a more compact notation.

\begin{align*}
  \begin{Bmatrix}
    \mathbf{X} \\
    \mathbf{Y} \\
    \mathbf{Z}
  \end {Bmatrix}
  &=
\begin{tikzpicture}[thick, baseline = {  ([yshift = -0.3em] current bounding box.center)  } ]
  \begin{pgfonlayer}{nodelayer}
    \node [style=tensor] (T0) at (0, -5em) {$\hat{\mathbf{T}}$};
    \node [style=tensor] (T1) at (8.0em, -5em) {$\hat{\mathbf{T}}$};
    \node [style=plain] (H0) at (4.0em, -3em) {$\mathbf{X}$};
    \node [style=plain] (H1) at (4.0em, -5em) {$\mathbf{Y}$};
    \node [style=plain] (H2) at (4.0em, -7em) {$\mathbf{Z}$};
  \end{pgfonlayer}
  \begin{pgfonlayer}{edgelayer}
	  \draw [very thick] (T0) to (H0.west);
	  \draw [very thick] (T0) to (H1.west);
	  \draw [very thick] (T0) to (H2.west);
	  \draw [very thick] (T1) to (H0.east);
	  \draw [very thick] (T1) to (H1.east);
	  \draw [very thick] (T1) to (H2.east);
  \end{pgfonlayer}
\end{tikzpicture}
\end{align*}

This has the properties of multilinearity and symmetry under permutation:
\begin{align*}
  %First, addition
  \begin{Bmatrix}
    \mathbf{W+X} \\
    \mathbf{Y} \\
    \mathbf{Z}
  \end {Bmatrix}
  & = %sum of two brackets
    \begin{Bmatrix}
    \mathbf{W}\\
    \mathbf{Y} \\
    \mathbf{Z}
  \end {Bmatrix} 
     + 
     \begin{Bmatrix}
    \mathbf{X} \\
    \mathbf{Y} \\
    \mathbf{Z}
  \end {Bmatrix} & \\ 
  %end of that equation, on to multiplication
     \begin{Bmatrix}
    \lambda \mathbf{X}\\
    \mathbf{Y} \\
    \mathbf{Z}
  \end {Bmatrix}
  & = 
    \lambda 
    \begin{Bmatrix}
    \mathbf{X} \\
    \mathbf{Y} \\
    \mathbf{Z}
  \end {Bmatrix} & \\
  %On to permutation
       \begin{Bmatrix}
    \mathbf{X}\\
    \mathbf{Y} \\
    \mathbf{Z}
  \end {Bmatrix}  & = 
    \begin{Bmatrix}
    \mathbf{Y} \\
    \mathbf{X} \\
    \mathbf{Z}
  \end {Bmatrix} 
  = 
    \begin{Bmatrix}
    \mathbf{Y} \\
    \mathbf{Z} \\
    \mathbf{X}
  \end {Bmatrix} 
\end{align*}

Furthermore, if \textit{all} the matrices are transposed, then you get the same value:
\begin{equation*}
    \begin{Bmatrix}
    \mathbf{X}\\
    \mathbf{Y} \\
    \mathbf{Z}
  \end {Bmatrix}  
  = 
  \begin{Bmatrix}
    \mathbf{X^T}\\
    \mathbf{Y^T} \\
    \mathbf{Z^T}
  \end {Bmatrix} 
\end{equation*}

Multilinearity and permutation symmetry mean that we can expand our bracket of sums in the same way as with a product of sums.

\begin{align*}
  %First, addition
	  \begin{Bmatrix}
	    \mathbf{\mathcal{A}} + \mathbf{\mathcal{A}^T} - \mathbf{I} \\
	    \mathbf{\mathcal{A}} + \mathbf{\mathcal{A}^T} - \mathbf{I}  \\
	    \mathbf{\mathcal{A}} + \mathbf{\mathcal{A}^T} - \mathbf{I} 
	  \end {Bmatrix}
	  & = %sum of two brackets
	    %coefficient
	   \begin{Bmatrix}
	    \mathbf{\mathcal{A}}\\
	    \mathbf{\mathcal{A}} \\
	    \mathbf{\mathcal{A}}
	  \end {Bmatrix} 
	     + 
	   \begin{Bmatrix}
	    \mathbf{\mathcal{A}^T} \\
	    \mathbf{\mathcal{A}^T} \\
	    \mathbf{\mathcal{A}^T}
	  \end {Bmatrix}  
	 \\ 
	  & \qquad + 
	  3    \begin{Bmatrix}
	    \mathbf{\mathcal{A}}\\
	    \mathbf{\mathcal{A}} \\
	    \mathbf{\mathcal{A}^T}
	  \end {Bmatrix} 
	  + 3    \begin{Bmatrix}
	    \mathbf{\mathcal{A}^T}\\
	    \mathbf{\mathcal{A}^T} \\
	    \mathbf{\mathcal{A}}
	  \end {Bmatrix} \\
	  & \qquad
	  - 3    \begin{Bmatrix}
	    \mathbf{\mathcal{A}}\\
	    \mathbf{\mathcal{A}} \\
	    \mathbf{I}
	  \end {Bmatrix} 
	    - 3    \begin{Bmatrix}
	    \mathbf{\mathcal{A}^T}\\
	    \mathbf{\mathcal{A}^T} \\
	    \mathbf{I}
	  \end {Bmatrix} 
	      - 6    \begin{Bmatrix}
	    \mathbf{\mathcal{A}}\\
	    \mathbf{I} \\
	    \mathbf{\mathcal{A}^T}
	  \end {Bmatrix} \\
	& \qquad
	  + 3    
	  \begin{Bmatrix}
	    \mathbf{\mathcal{A}}\\
	    \mathbf{I} \\
	    \mathbf{I}
	  \end {Bmatrix} 
	  + 3    
	  \begin{Bmatrix}
	    \mathbf{\mathcal{A}^T}\\
	    \mathbf{I} \\
	    \mathbf{I}
	  \end {Bmatrix} \\
	  & \qquad
	    -  \begin{Bmatrix}
	    \mathbf{I}\\
	    \mathbf{I} \\
	    \mathbf{I}
	  \end {Bmatrix} 
\end{align*}

Combining transposes lets us group the terms together further:

\begin{align*}
  %First, addition
	  \begin{Bmatrix}
	    \mathbf{\mathcal{A}} + \mathbf{\mathcal{A}^T} - \mathbf{I} \\
	    \mathbf{\mathcal{A}} + \mathbf{\mathcal{A}^T} - \mathbf{I}  \\
	    \mathbf{\mathcal{A}} + \mathbf{\mathcal{A}^T} - \mathbf{I} 
	  \end {Bmatrix}
	  & = %sum of two brackets
	    %coefficient
	   2 \begin{Bmatrix}
	    \mathbf{\mathcal{A}}\\
	    \mathbf{\mathcal{A}} \\
	    \mathbf{\mathcal{A}}
	  \end {Bmatrix} 
	 \\ 
	  & \qquad + 
	  6    \begin{Bmatrix}
	    \mathbf{\mathcal{A}}\\
	    \mathbf{\mathcal{A}} \\
	    \mathbf{\mathcal{A}^T}
	  \end {Bmatrix} 
	  - 6    \begin{Bmatrix}
	    \mathbf{\mathcal{A}}\\
	    \mathbf{\mathcal{A}} \\
	    \mathbf{I}
	  \end {Bmatrix} 
	  - 6    \begin{Bmatrix}
	    \mathbf{\mathcal{A}}\\
	    \mathbf{I} \\
	    \mathbf{\mathcal{A}^T}
	  \end {Bmatrix}
	  + 6
	  \begin{Bmatrix}
	    \mathbf{\mathcal{A}}\\
	    \mathbf{I} \\
	    \mathbf{I}
	  \end {Bmatrix} \\
	  & \qquad
	    -  \begin{Bmatrix}
	    \mathbf{I}\\
	    \mathbf{I} \\
	    \mathbf{I}
	  \end {Bmatrix} \\
	  & = 
	  2 \begin{Bmatrix}
	    \mathbf{\mathcal{A}}\\
	    \mathbf{\mathcal{A}} \\
	    \mathbf{\mathcal{A}}
	  \end {Bmatrix} 
	   + 
	  6    \begin{Bmatrix}
	    \mathbf{\mathcal{A}}\\
	    \mathbf{\mathcal{A}} \\
	    \mathbf{\mathcal{A}^T} - \mathbf{I}
	  \end {Bmatrix}
	  - 6    \begin{Bmatrix}
	    \mathbf{\mathcal{A}}\\
	     \mathbf{I} \\
	    \mathbf{\mathcal{A}^T} - \mathbf{I}
	  \end {Bmatrix} 
	    -  \begin{Bmatrix}
	    \mathbf{I}\\
	    \mathbf{I} \\
	    \mathbf{I}
	  \end {Bmatrix} \\
	  & = 
	  2 \begin{Bmatrix}
	    \mathbf{\mathcal{A}}\\
	    \mathbf{\mathcal{A}} \\
	    \mathbf{\mathcal{A}}
	  \end {Bmatrix}
	  + 6    \begin{Bmatrix}
	    \mathbf{\mathcal{A}}\\
	    \mathbf{\mathcal{A}} - \mathbf{I}\\
	    \mathbf{\mathcal{A}^T} - \mathbf{I}
	  \end {Bmatrix}
	    -  \begin{Bmatrix}
	    \mathbf{I}\\
	    \mathbf{I} \\
	    \mathbf{I}
	  \end {Bmatrix} 
\end{align*}
The calculation of $\begin{Bmatrix}
	    \mathbf{I}\\
	    \mathbf{I} \\
	    \mathbf{I}
	  \end {Bmatrix} $ is trivial: it's just $\langle \hat{\mathbf{T}}, \hat{\mathbf{T}} \rangle $. The other two involve grouping the tiny minority of \textit{nonzero} subterms in the first two terms. The principle to remember is that none of the nonzero levels are more than two apart in any of the 3 modes. This means that
$
	  \begin{Bmatrix}
	    \mathbf{\hat{A}^q} \\
	    \mathbf{ \hat{A} ^ r}  \\
	    \mathbf{ \hat{A} ^ s} 
	  \end {Bmatrix} = 0
$ if any of  
$\vert q-r \vert, \vert q-s \vert, \vert r-s \vert$ are more than 2, and
$
	  \begin{Bmatrix}
	    \mathbf{ \hat{A} ^ q} \\
	    \mathbf{ \hat{A} ^ r}  \\
	    \mathbf{ \hat{A} ^ {Ts}} 
	  \end {Bmatrix} = 0
$ 
if either $q+s$ or $r+s$ is greater than 2.
Taking the middle term first,

\begin{align*}
     \begin{Bmatrix}
	    \mathbf{\mathcal{A}} \\
	    \mathbf{\mathcal{A}} - \mathbf{I} \\
	    \mathbf{\mathcal{A}^T} - \mathbf{I}
     \end {Bmatrix}
     &=
      \begin{Bmatrix}
	   \mathbf{I} + \hat{\mathbf{A}} + \mathbf{ \hat{A}^2 } + ...\\
	   \hat{\mathbf{A}} + \mathbf{ \hat{A}^2 } + ...\\
	   \hat{\mathbf{A}} + \mathbf{\hat{A}^{T2}} + ...\\
     \end {Bmatrix} \\
          &=
      \begin{Bmatrix}
	   \mathbf{I} + \hat{\mathbf{A}}\\
	   \hat{\mathbf{A}}\\
	    \mathbf{\hat{A}^{T}}\
     \end {Bmatrix} \\
 \end{align*}

\begin{align*}
     \begin{Bmatrix}
	    \mathbf{\mathcal{A}}\\
	    \mathbf{\mathcal{A}} - \mathbf{I}\\
	    \mathbf{\mathcal{A}^T} - \mathbf{I}
     \end {Bmatrix}
     &=
      \begin{Bmatrix}
	    \sum_{q \ge 0} \mathbf{ \hat{A} ^ q}\\
	    \sum_{r > 0} \mathbf{ \hat{A} ^ r}\\
	   \sum_{s > 0} \mathbf{ \hat{A} ^ {Ts}} \\
     \end {Bmatrix} \\
     &=
     \sum_{q \ge 0} \sum_{r > 0} \sum_{s > 0} 
      \begin{Bmatrix}
	    \mathbf{ \hat{A} ^ q}\\
	    \mathbf{ \hat{A} ^ r}\\
	   \mathbf{ \hat{A} ^ {Ts}}
     \end {Bmatrix} \\ %which are zero if p > 1
     &=
     \sum_{s > 0} \sum_{q=0}^{2-s} \sum_{r=1}^{2-s}  
      \begin{Bmatrix}
	    \mathbf{ \hat{A} ^ q}\\
	    \mathbf{ \hat{A} ^ r}\\
	   \mathbf{ \hat{A} ^ {Ts}}
     \end {Bmatrix} 
\end{align*}
All terms have power $r, s \ge 1$. But this implies that the only nonzero terms have $q, r, s \le 1$. If $q$ were greater than 1, we would have $q+s > 1+s \ge 2$. $r \le 1$ for similar reasons. And if $s$ were greater than 1 we would have $s + r > 1 + r \ge 2$. Consequently, this triply-infinite sum reduces to

\begin{equation*}
     \begin{Bmatrix}
	    \mathbf{\mathcal{A}} \\
	    \mathbf{\mathcal{A}} - \mathbf{I} \\
	    \mathbf{\mathcal{A}^T} - \mathbf{I}
     \end {Bmatrix}
     =
     \sum_{q=0}^1 \sum_{r=1}^1 \sum_{s=1}^1
      \begin{Bmatrix}
	    \mathbf{ \hat{A} ^ q}\\
	    \mathbf{ \hat{A} ^ r}\\
	   \mathbf{ \hat{A} ^ {Ts}}
     \end {Bmatrix} \\
     = 
     \begin{Bmatrix}
	    \mathbf{I} + \hat{\mathbf{A}}\\
	   \hat{\mathbf{A}} \\
	   \mathbf{ \hat{A} ^ T}
     \end {Bmatrix}. \\
\end{equation*}

Finally, we reach the first sum, which is also the most complicated:

\begin{align*}
    \begin{Bmatrix}
	    \mathbf{\mathcal{A}}\\
	    \mathbf{\mathcal{A}} \\
	    \mathbf{\mathcal{A}}
    \end {Bmatrix}
    &= 
    \sum_{q \ge 0} \sum_{r \ge 0} \sum_{s \ge 0}
    \begin{Bmatrix}
	    \mathbf{ \hat{A} ^ q} \\
	    \mathbf{ \hat{A} ^ r}  \\
	    \mathbf{ \hat{A} ^ s} 
    \end {Bmatrix} \\
    &= 
    \sum_{q=r=s} 
    \begin{Bmatrix}
	    \mathbf{ \hat{A} ^ q} \\
	    \mathbf{ \hat{A} ^ r}  \\
	    \mathbf{ \hat{A} ^ s} 
       \end {Bmatrix}
     + 3 \sum_{q = r < s} 
         \begin{Bmatrix}
	    \mathbf{ \hat{A} ^ q} \\
	    \mathbf{ \hat{A} ^ r}  \\
	    \mathbf{ \hat{A} ^ s} 
       \end {Bmatrix}
    + 3 \sum_{q < r = s}
        \begin{Bmatrix}
	    \mathbf{ \hat{A} ^ q} \\
	    \mathbf{ \hat{A} ^ r}  \\
	    \mathbf{ \hat{A} ^ s} 
       \end {Bmatrix}
    + 6 \sum_{q < r < s} 
    \begin{Bmatrix}
	    \mathbf{ \hat{A} ^ q} \\
	    \mathbf{ \hat{A} ^ r}  \\
	    \mathbf{ \hat{A} ^ s} 
    \end {Bmatrix} \\
        &= 
    \sum_{q \ge 0} \begin{Bmatrix}
	    \mathbf{ \hat{A} ^ q} \\
	    \mathbf{ \hat{A} ^ q}  \\
	    \mathbf{ \hat{A} ^ q} 
    \end {Bmatrix} 
    + 3 \sum_{q \ge 0} %\sum_{\delta = 1}^2 %I prefer to make it explicit right now
    \Bigg(
        \begin{Bmatrix} 
          \mathbf{ \hat{A} ^ q} \\ 
          \mathbf{ \hat{A} ^ q}  \\
	    \mathbf{ \hat{A} ^ {q + 1} }
       \end {Bmatrix} 
       +         \begin{Bmatrix} 
          \mathbf{ \hat{A} ^ q} \\ 
          \mathbf{ \hat{A} ^ q}  \\
	    \mathbf{\hat{A}^{q + 2} }
       \end {Bmatrix} 
    \Bigg)
        + 3 \sum_{q \ge 0} %\sum_{\delta = 1}^2 
    \Bigg(
        \begin{Bmatrix}
	    \mathbf{ \hat{A} ^ q} \\
	    \mathbf{\hat{A} ^ {q + 1} }  \\
	    \mathbf{\hat{A} ^ {q + 1} }
    \end {Bmatrix} 
    +         \begin{Bmatrix}
	    \mathbf{ \hat{A} ^ q} \\
	    \mathbf{\hat{A}^{q + 2} }  \\
	    \mathbf{\hat{A}^{q + 2} }
    \end {Bmatrix} 
    \Bigg)
    + 6 \sum_{q \ge 0} 
    \begin{Bmatrix}
	    \mathbf{ \hat{A} ^ q} \\
	    \mathbf{\hat{A} ^ {q + 1} }  \\
	    \mathbf{\hat{A}^{q + 2} }
    \end {Bmatrix} \\
    &= 
     \sum_{q \ge 0} \begin{Bmatrix}
	    \mathbf{ \hat{A} ^ q} \\
	    \mathbf{ \hat{A} ^ q}  \\
	    \mathbf{ \hat{A} ^ q} 
       \end {Bmatrix} 
     + 3 \sum_{q \ge 0} %\sum_{\delta = 1}^2 %I prefer to make it explicit right now
        \begin{Bmatrix} 
          \mathbf{ \hat{A} ^ q} \\ 
          \mathbf{ \hat{A} ^ q}  \\
	  \mathbf{\hat{A} ^ {q + 1} } + \mathbf{\hat{A}^{q + 2} }
       \end {Bmatrix} 
     + 3 \sum_{q \ge 0} %\sum_{\delta = 1}^2 
        \begin{Bmatrix}
	    \mathbf{ \hat{A} ^ q} \\
	    \mathbf{\hat{A} ^ {q + 1} } + \mathbf{\hat{A}^{q + 2} } \\
	    \mathbf{\hat{A} ^ {q + 1} } + \mathbf{\hat{A}^{q + 2} }
    \end {Bmatrix} \\
    &= 
         \sum_{q \ge 0} \begin{Bmatrix}
	    \mathbf{ \hat{A} ^ q} \\
	    \mathbf{ \hat{A} ^ q}  \\
	    \mathbf{ \hat{A} ^ q} 
       \end {Bmatrix} 
     + 3 \sum_{q \ge 0} %\sum_{\delta = 1}^2 %I prefer to make it explicit right now
        \begin{Bmatrix} 
          \mathbf{ \hat{A} ^ q} \\ 
          \mathbf{ \hat{A} ^ q}  + \mathbf{\hat{A} ^ {q + 1} } + \mathbf{\hat{A}^{q + 2} } \\
	  \mathbf{\hat{A} ^ {q + 1} } + \mathbf{\hat{A}^{q + 2} }
	\end {Bmatrix} \\
	    &= 
         \sum_{q \ge 0} \begin{Bmatrix}
	    \mathbf{ \hat{A} ^ q} \\
	    \mathbf{ \hat{A} ^ q}  \\
	    \mathbf{ \hat{A} ^ q} 
       \end {Bmatrix} 
     + 3 \sum_{q \ge 0} %\sum_{\delta = 1}^2 %I prefer to make it explicit right now
        \begin{Bmatrix} 
          \mathbf{ \hat{A} ^ q} \\ 
          (\mathbf{I} + \mathbf{\hat{A}} + \mathbf{\hat{A}}^2) \mathbf{ \hat{A} ^ q} \\
	  (\mathbf{\hat{A}} + \mathbf{\hat{A}}^2 ) \mathbf{ \hat{A} ^ q}
	\end {Bmatrix}
\end{align*}

Using
\begin{equation*}
\mathcal{S} (\hat{\mathbf{T}}) =  \sum_{q \ge 0} \Big( \mathbf{ \hat{A} ^ q} \otimes \mathbf{ \hat{A} ^ q} \otimes \mathbf{ \hat{A} ^ q} \Big) (\hat{\mathbf{T}})
\end {equation*} 

then these terms become

\begin{align*}
         \sum_{q \ge 0} \begin{Bmatrix}
	    \mathbf{ \hat{A} ^ q} \\
	    \mathbf{ \hat{A} ^ q}  \\
	    \mathbf{ \hat{A} ^ q} 
       \end {Bmatrix} 
       &= \langle \hat{\mathbf{T}},  \mathcal{S} (\hat{\mathbf{T}}) \rangle \\
       \sum_{q \ge 0} 
        \begin{Bmatrix} 
          \mathbf{ \hat{A} ^ q} \\ 
          (\mathbf{I} + \hat{\mathbf{A}} + \mathbf{ \hat{A}^2 }) \mathbf{ \hat{A} ^ q} \\
	  (\hat{\mathbf{A}} + \mathbf{ \hat{A}^2 } ) \mathbf{ \hat{A} ^ q}
	\end {Bmatrix}
	&= \langle \hat{\mathbf{T}},  
	            \Big( \mathbf{I} \otimes (\mathbf{I} + \hat{\mathbf{A}} + \mathbf{ \hat{A}^2 }) \otimes (\hat{\mathbf{A}} + \mathbf{ \hat{A}^2 }) \Big) (\mathcal{S} (\hat{\mathbf{T}}))
	            \rangle 
\end{align*}

%FINALLY, WRAPPING UP THE PROOF OF THE EXPRESSION,

Putting the whole expression together,

\begin{align*}
IIIb 
  & =   \begin{Bmatrix}
	    \mathbf{\mathcal{A}} + \mathbf{\mathcal{A}^T} - \mathbf{I} \\
	    \mathbf{\mathcal{A}} + \mathbf{\mathcal{A}^T} - \mathbf{I}  \\
	    \mathbf{\mathcal{A}} + \mathbf{\mathcal{A}^T} - \mathbf{I} 
	  \end {Bmatrix} \\
  & = 	   2 \begin{Bmatrix}
	    \mathbf{\mathcal{A}}\\
	    \mathbf{\mathcal{A}} \\
	    \mathbf{\mathcal{A}}
	  \end {Bmatrix} 
  + 6 \begin{Bmatrix}
	    \mathbf{\mathcal{A}} \\
	    \mathbf{\mathcal{A}} - \mathbf{I} \\
	    \mathbf{\mathcal{A}^T} - \mathbf{I}
         \end {Bmatrix} 
   -  \begin{Bmatrix}
	    \mathbf{I} \\
	    \mathbf{I} \\
	    \mathbf{I}
         \end {Bmatrix} 
         \\
     &= 2 \Big( \langle \hat{\mathbf{T}}, \mathcal{S} (\hat{\mathbf{T}}) \rangle + 3 \langle \hat{\mathbf{T}},\Big( \mathbf{I} \otimes (\mathbf{I} + \hat{\mathbf{A}} + \mathbf{ \hat{A}^2 }) \otimes (\hat{\mathbf{A}} + \mathbf{ \hat{A}^2 }) \Big) (\mathcal{S} (\hat{\mathbf{T}})) \rangle \Big)
     + 6   \begin{Bmatrix}
	      \mathbf{I} + \hat{\mathbf{A}}\\
	      \hat{\mathbf{A}} \\
	      \mathbf{ \hat{A} ^ T}
            \end {Bmatrix}
     -  \begin{Bmatrix}
	    \mathbf{I} \\
	    \mathbf{I} \\
	    \mathbf{I}
         \end {Bmatrix} 
\end{align*}

%MAY WELL WANT TO SUBSTITUTE S(T) FOR Z!!

Like the computation in {near-diags} there is a simple, linear recurrence relation for $\mathcal{S} (\hat{\mathbf{T}})$. First, $\mathcal{S} (\hat{\mathbf{T}})$ solves a linear equation:

\begin{align*}
\mathcal{S} (\hat{\mathbf{T}}) &= \hat{\mathbf{T}} + \sum_{q > 0} \Big( \mathbf{ \hat{A} ^ q} \otimes \mathbf{ \hat{A} ^ q} \otimes \mathbf{ \hat{A} ^ q} \Big) (\hat{\mathbf{T}}) \\
&= \hat{\mathbf{T}} + \Big( \hat{\mathbf{A}} \otimes \hat{\mathbf{A}} \otimes \hat{\mathbf{A}} \Big) (\mathcal{S} (\hat{\mathbf{T}}))
\end{align*}

In block notation, this gives us a recurrence relation similar to \ref{Solving S}. This is a forward recurrence instead of backwards because of 

\begin{align*}
\mathcal{S} (\hat{\mathbf{T}})_{ijk} &= \hat{\mathbf{T}}_{ijk} + \sum_{i' j' k'} \Big( \hat{\mathbf{A}}_{ii'} \otimes \hat{\mathbf{A}}_{jj'} \otimes \hat{\mathbf{A}}_{kk'} \Big) (\mathcal{S} (\hat{\mathbf{T}})_{i' j' k'}) \\
&= \hat{\mathbf{T}}_{ijk} \text{ if i=1 or j=1 or k=1} \\
&=  \hat{\mathbf{T}}_{ijk} + \Big( \hat{\mathbf{A}}_{i, i-1} \otimes \hat{\mathbf{A}}_{j, j-1} \otimes \hat{\mathbf{A}}_{k, k-1} \Big) (\mathcal{S} (\hat{\mathbf{T}})_{i-1, j-1, k-1}) \ \text{otherwise}
\end{align*}
%if $min(i, j, k) > 1$, otherwise simply $\hat{\mathbf{T}}_{ijk}$.
%In other words, 

The recurrence is forward instead of backward because we chose to use $\sum_{q > 0} \Big( \mathbf{ \hat{A} ^ q} \otimes \mathbf{ \hat{A} ^ q} \otimes \mathbf{ \hat{A} ^ q} \Big) (\hat{\mathbf{T}})$ rather than  $\sum_{q > 0} \Big( \mathbf{ \hat{A} ^ {Tq}} \otimes \mathbf{ \hat{A} ^ {Tq}} \otimes \mathbf{ \hat{A} ^ {Tq}} \Big) (\hat{\mathbf{T}})$ for the main sum. Each successive block of $\mathcal{S} (\hat{\mathbf{T}})$ is computed in terms of a known block of $\hat{\mathbf{T}}$ and (if present) the previous block of $\mathcal{S} (\hat{\mathbf{T}})$ from the same level. There are 7 block levels, the number of blocks per level is $O(n)$, and the number of operations in a matrix-times-block operation is $O(p^4)$, so the complexity of finding $\mathcal{S} (\hat{\mathbf{T}})$ is $O(np^4)$.

The computation can be accelerated by separating the levels of $\mathbf{\hat{T}}$ and making greater use of symmetry, but it will still be $O(np^4)$.

\bibliographystyle{plainnat} % $\mathbf{H}OME/latex/asa}
\bibliography{Sparse-Laplace}   

\end{document}